# STATISTICAL MODELS FOR AVERAGING OF THE PUMP-PROBE TIME TRACES: EXAMPLE OF DENOISING IN TERAHERTZ TIME DOMAIN SPECTROSCOPY

M. Skorobogatiy*, J. Sadasivan, H. Guerboukha

Engineering Physics, Polytechnique de Montréal, C.P. 6079, succ. Centre-ville, Montréal (QC) Canada, H3C 3A7
* - corresponding author; maksim.skorobogatiy@polymtl.ca

Matlab code implementing statistical models detailed in this paper is available at:

www.polymtl.ca/phys/photonics/PolyFIT_THz/PolyFIT_THz_code.zip

**Abstract.** In this paper, we first discuss the main types of noise in a typical pump-probe system, and then focus specifically on terahertz time domain spectroscopy (THz-TDS) setups. We then introduce four statistical models for the noisy pulses obtained in such systems, and detail rigorous mathematical algorithms to de-noise such traces, find the proper averages and characterise various types of experimental noise. Finally, we perform a comparative analysis of the performance, advantages and limitations of the algorithms by testing them on the experimental data collected using a particular THz-TDS system available in our laboratories. We conclude that using advanced statistical models for trace averaging results in the fitting errors that are significantly smaller than those obtained when only a simple statistical average is used.

## 1. Introduction

Pump-probe technique has received a widespread use in a number of physical and engineering fields including, ultrafast spectroscopy of chemical reactions [1], nanomaterials [2], semiconductor materials [3], time resolved microscopy [4-6], time-resolved Magneto-Optic Kerr effect (MOKE) [7] and most recently in THz spectroscopy. Within this technique, a laser beam (pump) is used to excite a sample, and a time delayed laser beam (probe) interacts with the excited sample, thus probing it at various precisely controlled temporal moments. The pump and the probe can be from the same source or from two different sources. A net result of the pump-probe technique is a collection of the temporary resolved pulses that represent dynamics of a studied physical process. While all the pulses contain identical physical information, each one of them also includes a noise contribution that is unique for each pulse. The question then is how to properly average the measured time traces and their spectra in order to extract the physical information and mitigate noise. A measure of effectiveness of a given statistical approach can be, for example, the value of a Signal to Noise Ratio (SNR).

## 2. Standard treatment of the pump-probe data

There have been various methods, both mathematical and experimental to characterize and optimize the noise and enhance the Signal to Noise Ratio for measurements obtained from a pulse-probe experiment. For example, in [8], averaging of the signal was done using two different methods - slow scanning (modulated pump / synchronous probe demodulation) and fast scanning (unmodulated pump) on a semiconductor film and a mathematical model was developed to analytically describe the SNR of the measurements. Research has also been conducted into identification and classification of noise patterns in kilohertz frequency pump-probe experiments [9]. There are also many instances for characterization and optimization of noise levels in a THz time domain pump-probe setup. Mira Naftaly et al. in [10], have shown the effects of noise in measurements using a THz TDS system and suggested some practical measures in calibrating the experimental set-up and eliminating the noise. In [11], analysis of noise and uncertainties in THz TDS setup has been done with keeping material parameter extraction in mind. Effects of different types of noise affecting the PCA antennas including the laser fluctuations, thermal noise etc. has been discussed in [12] along with practical measures to mitigate them. De-noising algorithm involving wavelet transforms have also been used to reduce the noise levels of measurements from the THz TDS set-ups which can enhance the SNR of the measurements [13, 14]. Most of the statistical methods have been developed and verified only for specific applications, for example in [15], digital signalling process have been utilised to de-noise the data obtained from a THz pulsed imaging system for biological samples. It has been shown that the SNR can also be enhanced by using the Terahertz Differential Time Domain Spectroscopy (DTDS) technique rather than the normal method for liquids [16], thin films [17, 18]. In Terahertz TDS setup which employs electro-optic crystals for generation and detection of THz beams, it has been shown that SNR can be enhanced by reducing distortion in the system by using quarter wave plate to increase the initial birefringence of the probe beam [19].

The goal of this paper is to revisit some of the common types of noise encountered in pump-probe experiments, and try to counter them with general mathematical models that are non-specific to physical nature of the studied systems. Surprisingly, such analysis has not been reported so far, to our knowledge. As we show in the following, our analysis not only allows dramatic enhancement in the SNR of the pump probe measurements, but it also allows monitoring slow time variations in the key experimental parameters such as power, phase and jitter. Before we get into details of our approach we remind the reader the basics of statistical analysis of the pump-probe data.

The most straightforward way of treating a collection of the pump-probe pulses is to take a simple average of the pulse spectra $E_p(\omega)$ assuming that the noise contribution is a trace-wise random variable with a zero mean and a standard deviation that decreases with an increasing number of traces. Then, for the spectra of a nominal pulse $E_{sa}(\omega)$ ($sa$ stands for "simple average") that describes the physical process studied by a pulse-probe technique, spectrally dependent standard deviation of a noise $\delta E_{sa}^2(\omega)$, and a corresponding frequency dependent Signal to Noise Ratio $SNR_{sa}(\omega)$ we write:

$$E_{sa}(\omega) = \frac{1}{N_t}\sum_{p=1}^{N_t} E_p(\omega), \quad (1)$$

$$\delta E_{sa}^2(\omega) = \frac{1}{N_t}\sum_{p=1}^{N_t}\left|E_p(\omega) - E_{sa}(\omega)\right|^2, \quad (2)$$

$$SNR_{sa}(\omega) = \frac{E_{sa}(\omega)}{\sqrt{\delta E_{sa}^2(\omega)}}, \quad (3)$$

where $N_t$ is the number of experimental traces. We note that expressions (1) and (2) can be obtained by solving a certain minimisation problem. Indeed, assuming that the spectra of the individual traces are related to that of a nominal pulse $E_{sa}(\omega)$ and a trace-dependent noise $\delta E_p(\omega)$ as:

$$E_p(\omega) = E_{sa}(\omega) + \delta E_p(\omega), \quad (4)$$

expressions (2.1), (2.2) can be obtained by minimizing the following weighting function with respect to the spectrum of a nominal pulse:

$$Q = \frac{1}{N_t}\sum_{p=1}^{N_t}\frac{1}{N_\omega}\sum_{n=1}^{N_\omega}\left|\delta E_p(\omega_n)\right|^2 = \frac{1}{N_t}\sum_{p=1}^{N_t}\frac{1}{N_\omega}\sum_{n=1}^{N_\omega}\left|E_p(\omega_n) - E_{sa}(\omega_n)\right|^2, \quad (5)$$

where $N_\omega$ is the number of frequencies in the spectra of experimental pulses, which are computed using discrete Fourier transform. From (2) it also follows that the value of $Q$ corresponds to the spectral average of noise $Q = \frac{1}{N_\omega}\sum_{n=1}^{N_\omega}\delta E_{sa}^2(\omega_n)$. In practical terms, minimization of $Q$ with respect to the complex spectra of the nominal pulse $E_{sa}(\omega)$ entails solution of the following system of $N_\omega$ equations:

$$\min_{E_o} Q \Leftrightarrow \frac{\partial Q}{\partial E_{sa}^*(\omega_n)} = 0; \ n = 1 \ldots N_\omega. \quad (6)$$

The main disadvantage of this algorithm and the underlying analytical fitting model (4) is an assumption that the main source of noise is in the form of an additive linear contribution to the nominal signal, and that such contribution has a zero trace-wise average. In fact, one has to recognize that there are several sources of noise in a typical pump-probe system. Moreover, these sources of noise do not contribute in a linear additive fashion, and, therefore, cannot be accounted for using simple fitting models like (4).

The purpose of this paper is first to establish the main types of noise contribution in a typical pump-probe system, then introduce several realistic fitting models for the pulses obtained in such systems, as well as mathematical algorithms to find the fitting parameters, and finally investigate performance of such algorithms and their limitations. To make our discussion more concrete, we will use the data from a standard Terahertz Time Domain Spectroscopy (TDS-THz) system (see in Fig. 1), which is a particular realisation of a pump-probe experiment.

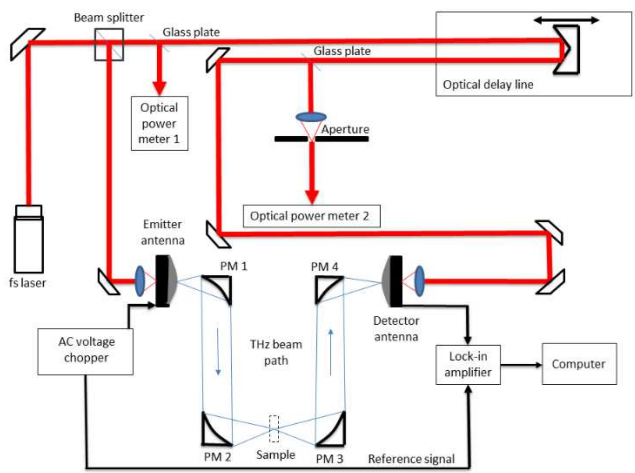

Fig. 1: Schematic of the Terahertz Time Domain Spectroscopy (THz- TDS) setup.

## 3. Sources of noise in a typical pump-probe system

In order to develop more advanced fitting models for the data obtained in a pump-probe experiment we have to establish the main sources of noise in such systems. Consider for example, a TDS-THz system shown in Fig. 1, which is a particular implementation of a pump-probe system. There, an ultrafast femtosecond laser emits a Near-Infrared (NIR) beam, which is furthermore divided into two beams using a 50:50 beam-splitter. The first NIR beam is then incident onto a photoconductive emitter antenna.  An AC voltage source is connected to the emitter antenna acting as an electronic modulator of the emitted THz beam. In other implementations, a DC voltage source is used with an emitter antenna, while a mechanical chopper is employed to interrupt the NIR beam, and thus modulate the emitted THz beam. The generated THz beam is then collimated and focused using a pair of parabolic mirrors. After interacting with a sample, the THz beam is focused onto a photoconductive detector antenna. The second NIR beam is used to excite the detector antenna after passing through a variable optical delay line. The optical delay line allows the THz pulse to be measured as a function of time by delaying the second NIR beam which gates the detector antenna. Furthermore, the THz pulse is recorded using a lock-in amplifier whose reference frequency is set by the AC voltage source (or a mechanical chopper) that modulate the emitter antenna. Finally, we also monitor the time variation of the laser power, as well as variation of the optical power after the delay line by using thin glass plates to divert a small portion of the laser beam into optical power meters 1 and 2. These two measurements are useful in order to assure proper laser operation and proper optical alignment of various optical components during the measurements.

A result of a single compete measurement using a TDS-THz system with (or without) a sample inside is a temporary resolved transmitted THz pulse. In fact, when performing a careful spectroscopic study, in addition to using a relatively large averaging time constant (10ms-1000ms) of a lock-in amplifier, one also acquires a relatively large number (10-1000) of pulses to further mitigate the noise. In fact, the use of a lock-in amplifier and proper averaging of the time traces serve two different purposes. Thus, an analogue averaging performed by a lock-in amplifier is meant to reduce the contribution from a fast-varying noise coming mostly from various electronic and opto-electronic components, therefore larger values of a time constant are desired to reduce such a noise. At the same time, a complete pulse measurement using lock-in amplifier has to be fast enough compared to the time scales set by other slow varying processes in a system like antenna aging, laser power variation, or changes in the environmental conditions like humidity and temperature that happen on a scale of minutes to hours. Therefore, the lock-in time constant cannot be too large. Ideally, with the proper choice of a lock-in time averaging constant, individual time pulses should feature only a small contribution from the fast variable electronic noise, while they could be considered as acquired at constant environmental and experimental conditions.

In order to mitigate the effects of a slow-varying noise caused by the drift of various experimental parameters mentioned above, one typically resorts to acquiring a relatively large number of pulses that should be properly averaged to extract a desired physical information. Due to the fact that over time of a standard experiment (hours-days), time average of many slow varying parameters is not zero, simple averaging of traces as given by (1) will not be efficient in mitigating this type of "slow" noise. Therefore, system-specific fitting functions have to be introduced in order to compensate for the slow-varying trace-to-trace drifts of various experimental parameters.

We now investigate in more details the pump-probe system presented in Fig. 1 and identify the principle sources of noise in such systems. First, we note that the average power of an fs laser varies over time. A time resolved measurement using optical meter of a Menlo C-Fiber 780 Femtosecond Erbium laser shows ~ 1% variation on a 60 min time scale, which could affect the THz pulse amplitude. Additionally, the photoconductive antennas used for THz generation age over time resulting in lower emitted power under the same excitation conditions, which will again affect the THz pulse amplitude. In our case, we estimate ~10% variation in the antenna gain over 24 hours of continuous use. Moreover, when measuring THz pulse transmission under ambient not temperature stabilised conditions, humidity variation causes changes in the THz absorption (pulse amplitude), while temperature variation causes changes in the pulse phase and propagation time due to drifts in the setup physical dimensions, as well as refractive indices of various optical elements. The time scale for such variations in the environmental conditions in our lab is ~1 hour. Finally, we note that the optical delay line used in our experiments introduces an addition trace-to-trace variation in the pulse position. This is related to the fact that a typical optical delay line uses a retroreflector mirror mounted onto a linear micro-positioning stage that show variation and drift in its absolute position during repetitive use. Particularly, in our lab we use a Newport delay line that is specked for 1.5 $\mu m$ (unidirectional) precision repeatability in the absolute position of the delay line, which amounts to 0.01ps trace-to-trace variation in the position of the pulse. As an example, in Fig. 2 we plot result of a typical TDS-THz measurement of an empty system. We present ten consecutively acquired THz pulses and notice three major types of trace-to-trace variations which are pulse amplitude, pulse phase, and pulse position (pulse jitter) variations.

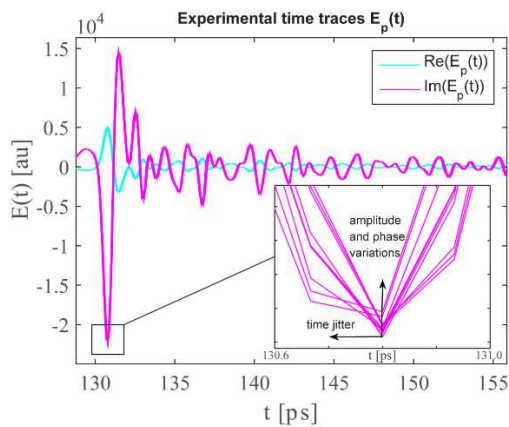

Figure 2. Example of the experimental THz traces with various types of noise identified.

In what follows we present three different analytical models that take into account the three abovementioned types of "slow" noise that include pulse power and phase variations, as well as pulse jitter. The three models are different in the way they include the leftover "fast" noise which is not completely compensated by the lock-in amplifier. We then detail mathematical formulation for fitting experimental data using the proposed analytical models. Finally, we perform a comparative analysis for the quality of the extracted data by comparing results of simple averaging versus more advanced algorithms discussed in this paper.

## 4. Advanced models to mitigate noise in the pump-probe experiments

Let us denote the nominal THz field emitted by the photoconductive antenna as $E_o(t)$. Due to various sources of noise, like laser power and phase variation over time, imprecision in the absolute position of the optical delay line from one run to another, as well as noise incurred during beam propagation and emission / detection, the THz field of trace $p$ detected by the receiver antenna $E_p(t)$, as well as its spectrum $E_p(\omega)$ can be written in the simplest form as:

Model 1:

$$E_p(t) = C_p E_o(t - \delta t_p) + \delta E_p(t - \delta t_p)$$

$$E_p(\omega) = \left( C_p E_o(\omega) + \delta E_p(\omega) \right) e^{-i\omega_n \delta t_p}, \quad (7)$$

In the expression above, $C_p$ are the complex gain factors of the transmission system $C_p \sim E_p/E_o$ that account for "slow" changes in the system optical properties happening between acquisition of the two consecutive traces, and that cannot be removed by a lock-in amplifier. Those include: laser power variation, emitter antenna aging, changing optical path absorption and optical path phase variation. Additionally, we introduce $\delta t_p$, which are the temporal shifts of the THz pulses due to imprecision in the absolute position of the optical delay line from one scan to another. Finally, the leftover "fast" noise which is not completely compensated by the lock-in amplifier is denoted as $\delta E_p(t)$. This noise is incurred either during pulse propagation or emission/detection and it varies on a time scale which is much faster than the time of a single measurement. Note that in model (7) we assume that the noise $\delta E_p(t)$ is independent of the pulse amplitude $C_p$. This is a valid assumption if the noise is purely electrical in nature and comes, for example, from the semiconductor carrier density fluctuation in the photoconductive antenna, as well as other processes that are either independent of the photoexcitation process or that happen while operating in the saturation regime of the photoexcitation process. Here by saturation regime we mean the case when increasing laser power does not lead to increase in the THz signal intensity.

Alternatively, if the "fast" noise is generated while operating in the unsaturated regime of the photoexcitation process during pulse generation or detection (for example photoexcited carrier density fluctuations), then it is reasonable to assume that the noise should be proportional to the amplitude of the excitation laser beam. We can then write a new model for the relation between the nominal and the registered traces as:

Model 2:

$$E_p(t) = C_p \left( E_o(t - \delta t_p) + \delta E_p(t - \delta t_p) \right)$$

$$E_p(\omega) = C_p\left(E_o(\omega) + \delta E_p(\omega)\right)e^{-i\omega_n\delta t_p}. \quad (8)$$

Finally, if the "fast" noise is incurred during pulse propagation due to rapidly changing environmental factors like variable air flows, sudden humidity variations, or changing air particulate density, then it is reasonable to assume that the noise should be proportional to the amplitude of the pulse itself, and we can write a third model for the relation between the nominal and the registered traces as:

Model 3:

$$E_p(t) = C_p E_o(t - \delta t_p)\left(1 + \delta_p(t - \delta t_p)\right)$$

$$E_p(\omega) = C_p E_o(\omega)e^{-i\omega\delta t_p}\left(1 + \delta_p(\omega)\right). \quad (9)$$

Ideally, given an experimental data, one should be able to infer the nature of the "fast" noise by performing comparative analysis of the data using models 1-3. Particularly, if the "fast" noise is generated by the non-saturated photoexcitation process or by the rapidly changing environmental factors, then models 2,3 should result in the noise amplitudes $\delta E_p, \delta_p$ that are independent of the gain factors $C_p$. At the same time, model 1 would predict noise amplitudes which are proportional to the gain factors $\delta E_p \sim C_p$. Alternatively, if the "fast" noise is generated by the saturated photoexcitation process or by electronic processes that are independent of the laser power and pulse amplitude, then model 1 should predict noise amplitudes $\delta E_p$ which are independent of the gain factors $C_p$, while models 2,3 should predict noise amplitudes which are inversely proportional to the gain factors $\delta E_p, \delta_p \sim 1/C_p$.

Finally, we note that although the three models differ only in the analytical representation of the "fast" noise, their mathematical treatments will be quite different from each other as we will see in what follows. In all three cases, we will find the fitting parameters by using minimization of the corresponding weighting functions that are defined to be proportional to the average value of the "fast" noise in the data.

## 5. Finding fitting parameters by solving optimization problem

In order to find various fitting parameters used in the analytical forms of the fitting functions (7-9), we first define model-dependent "fast" noise as:

Model 1: $\delta E_p(\omega) = E_p(\omega)e^{i\omega_n\delta t_p} - C_p E_o(\omega), \quad (10)$

Model 2: $\delta E_p(\omega) = a_p E_p(\omega)e^{i\omega_n\delta t_p} - E_o(\omega)\,;\; a_p = \frac{1}{C_p}, \quad (11)$

Model 3: $\delta_p(\omega) = ln\left(E_p(\omega)\right) - ln\left(C_p\right) - ln\left(E_o(\omega)\right) + i\omega\delta t_p$ ; $\delta_p(\omega) \ll 1$.   (12)

In order to find the complex gain factors $C_p$ (or their inverse $a_p$), time shifts $\delta t_p$, and a spectrum of the nominal THz field $E_o(\omega)$ we define an optimization problem with respect to the spectrally and trace-wise averaged value of the noise. In other words, we look for the values of the abovementioned fitting parameters that minimize the following weighting functions Q:

Models 1,2: $Q = \frac{1}{N_t}\sum_{p=1}^{N_t}\frac{1}{N_\omega}\sum_{n=1}^{N_\omega}\left|\delta E_p(\omega_n)\right|^2$,    (13)

Model 3: $Q = \frac{1}{N_t}\sum_{p=1}^{N_t}\frac{1}{N_\omega}\sum_{n=1}^{N_\omega}\left|\delta_p(\omega_n)\right|^2$,    (14)

where $N_t$ is the number of THz pulses used in the fitting, and $N_\omega$ is the number of frequency components in the Fourier spectrum of the measured pulses. In order to find the minimum of the weighting functions (13,14) we have to solve a system of $N_\omega + 2N_t$ generally non-linear equations:

Models 1,2: $\frac{\partial Q}{\partial E_o^*(\omega_n)} = 0$ ; Model 3: $\frac{\partial Q}{\partial \ln(E_o(\omega_n))^*} = 0$ ; $n = 1 \dots N_\omega$,    (15)

Model 1: $\frac{\partial Q}{\partial c_p^*} = 0$ ; Model 2: $\frac{\partial Q}{\partial a_p^*} = 0$ ; Model 3: $\frac{\partial Q}{\partial \ln(C_p)^*}$ ; $p = 1 \dots N_t$,    (16)

Models 1,2,3: $\frac{\partial Q}{\partial \delta t_p} = 0$; $p = 1 \dots N_t$.    (17)

Additionally, we note that for each model, a system of equations (15-17) is ill-posed and does not lead to a unique solution. This is related to the fact that the analytical fitting functions (7-9) are degenerate with respect to certain transformations of the groups of the fitting variables as explained further in the text. Therefore, in order to obtain a unique solution to (15-17) we need to introduce additional constraints which come in the form of various normalization conditions for the complex gain coefficients, as well as an assumption about the mean value of the time shifts. Furthermore, a much involved discussion is necessary in the case of the model 3 due to the fact that natural logarithm of a complex function gives a complex phase defined up to an unknown $2\pi$ multiplier, thus requiring careful phase unwrapping and removal of various numerical artefacts.

Finally, we note that in the case of models 1,2 a system of equations (15-17) contains $N_\omega + N_t$ linear equations (15,16), as well as $N_t$ nonlinear equations (17). Therefore, solution of such equations requires an iterative method which can be computationally intensive when dealing with a large number of THz traces. That said, we find that non-linear system of equations (15-17) is very stable and solution can be found readily after only a few iterations of any standard iterative algorithm, like a Newton method. At the same time, model 3 results in a system of $N_\omega + 2N_t$ linear equations that can be furthermore solved in a closed form. Therefore, model 3 is considerably less numerically intensive than models 1,2. That said, we find that model 3 is more unstable and it is more prone to noise than models 1,2, which is a direct consequence of the need to properly unwrap phases of all the traces before using model 3. We therefore conclude that models 1,2 are preferred when analysing sets of pump-probe traces as they give highly stable and predictable results over all frequency regions covered by the pulse spectra. At the same time, model 3 is preferred when one needs direct access to a continuous phase of the fitted nominal trace, which is necessary when extracting effective refractive index of the propagation media, or when interpreting results of the cut-back measurements of waveguides.

Due to length and complexity of the mathematical formulation of the three algorithms we have placed them in the Appendix section of the paper. In the following sections, we rather concentrate on the comparative analysis between the three models, while referring the reader to Appendix section for mathematical details related to the fitting algorithm implementation.

## 6. Application of models 1,2 to analysis of the time pulses

In this section, we present an example of analysis of a collection of 400 THz time traces acquired using an in-house THz-TDS system. We note that the following analysis aims only at giving an example of statistical treatment of the THz traces, while involving only a single set of pulses. Therefore, the conclusions of this analysis are not representative of all the different THz-TDS systems, neither of the different excitation and detection regimes that can be realised using such systems. We stress that in this paper we do not aim at answering the question about which model 1,2 or 3 and in which operation regime is more applicable to represent noise in the THz-TDS systems. The aim of this paper is, rather, to introduce several plausible models for the modelling and negation of the different types of noise, while the aim of this section is to simply demonstrate the type of results that can come from such an analysis.

The experimental system features a frequency doubled Menlo C-Fiber 780 Femtosecond fiber-laser (90 fs, repetition rate of 100 MHz, wavelength of 780 nm), and the photoconductive antennas deposited on low-temperature-grown GaAs substrates (TERA8-1, Menlo Systems). NIR optical beams of ~10 mW power were used to excite and gate the emitter and detector antennas, while a 20 V, 24 kHz AC voltage source was used to modulate the emitter antenna and a THz signal. The THz pulses were acquired using the SR830 Lock-in Amplifier (Stanford Research Systems). Experimental parameters during pulse acquisition were chosen as follows: 10 ms lock-in time constant of the amplifier, 20 $\mu m$ step size of the delay line (corresponding time resolution of 0.133 ps, and a maximal frequency of 3.75 THz), a total optical delay of 100 mm (corresponding frequency resolution of 3 GHz). The acquisition time of a single THz pulse was 4 minutes. The delay line precision for the absolute position reproducibility was 1.5 $\mu m$ (unidirectional), thus resulting in time jitter of pulses ~0.01 ps. Additionally, we have used a Newport 841 PE optical power meter to record the laser intensity before each trace measurement, as well as to record the NIR power after the delay line (see Fig. 1). During experiment, we measure both real $E_x(t)$ and imaginary $E_y(t)$ parts of the THz pulses, which are read at the "X" and "Y" output channels of a lock-in amplifier. The total complex measured THz trace is then obtained using $E(t) = E_x(t) + iE_y(t)$.

In Fig. 3 we show real (solid cyan) and imaginary parts (solid magenta) of the experimentally acquired traces $E_p(t)$, as well as real (dotted blue) and imaginary (dotted red) parts of the fitted nominal pulse $E_o(t)$ that were found using model 1 (7), and by minimizing weighting function (13). Visually, there is an excellent agreement between the experimental measurements and a theoretical fit. At the same time, we notice that some experimental traces deviate significantly from an ensemble of other traces, and as we will see in the following, they are characterized by strong variation of some (or all) of their key parameters (power, phase, time shift) from the rest of the traces. This normally happens due to some unforeseen significant fluctuations in the system (like a stuck delay stage, sudden draft of air when a person passes next to a measurement setup, etc.), from which the system rapidly recovers on the following trace acquisition. It is normally a good idea to remove such traces from the further analysis as they are not representative of the trace ensemble.

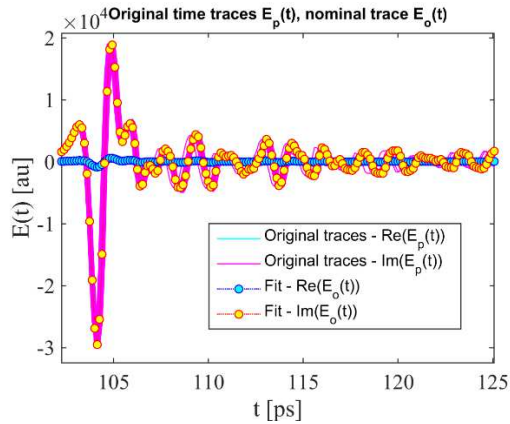

Figure 3. Experimentally measured time traces $E_p(t)$ (solid curves) and the nominal pulse fit $E_o(t)$ (dotted curves with circles).

In Fig. 4 (left panel) we plot spectra $E_p(\omega)$ of the experimental pulses (solid green), spectrum of the fitted nominal pulse $E_o(\omega)$ (solid blue), and a spectrum of the fitting error $\delta E_o(\omega)$ (solid red) defined as a trace-wise average of the individual fitting errors (10):

$$\delta E_o^2(\omega) = \frac{1}{N_t} \sum_{p=1}^{N_t} \left| \delta E_p(\omega) \right|^2. \quad (18)$$

Additionally, in Fig. 4 (right panel) we present a frequency dependent signal to noise ratio $SNR_o(\omega)$ for the model 1 fit defined as:

$$SNR_o(\omega) = \frac{E_o(\omega)}{\sqrt{\delta E_o^2(\omega)}}, \quad (19)$$

For comparison, in Fig. 4 we also present spectrum of the nominal pulse $E_{sa}(\omega)$ (dashed cyan) as calculated using simple averaging Eq. (1), an associated fitting error $\delta E_{sa}(\omega)$ (dashed magenta) as calculated using Eq. (2), as well as a frequency dependent signal to noise ratio $SNR_{sa}(\omega)$ computed for the simple average approximation according to (3).

From Fig. 4 we can see clearly that using advanced fitting model 1 results in a considerably more precise fit of the experimental data compared to the simple average approximation. Indeed, the spectrum of the fitting error $\delta E_o(\omega)$ for model 1 features values that vary in a relatively narrow range of $2 - 5 \cdot 10^3$ $au$ throughout the whole frequency range (see Fig.4 (left panel)). At the same time, the values of the fitting error $\delta E_{sa}(\omega)$ for a simple average approximation are strongly frequency dependent and vary in a much larger range of $2 - 30 \cdot 10^3$ $au$. Similarly, signal to noise ratio for the model 1 is considerably higher than that for the simple average approximation at most frequencies; thus, $SNR_o(\omega)$ reaches the value of $\sim 75$ for model 1, while the maximum $SNR_{sa}(\omega)$ for a simple average approximation is only $\sim 15$.

From Fig. 4 we also notice that in the near vicinity of the water vapor absorption lines, both the fitting error and SNR as obtained using model 1 does not show any improvements compared to those computed using simple averaging. This is related to the fact that in these spectral regions, model 1 does not capture the nature of the noise origin. In fact, in order to work properly near the water absorption lines, model 1 has to be further augmented to account for the trace-to-trace variation in the optical path absorption that can be caused, for example, by small changes in the optical path length. Although not explored in this paper, model 1 can be further improved by introducing, for example, a $e^{-i\alpha(\omega_n)\delta L_p}$ multiplier in Eq. (7), where $\alpha(\omega_n)$ is an absorption loss of water vapor during experiment, while $\delta L_p$ are new fitting parameters that characterize trace-to-trace variation in the optical path length. As most of the THz-TDS experiments are conducted in the dry nitrogen environment, or the spectral data is used away from the water absorption lines, we did not pursue this issue further.

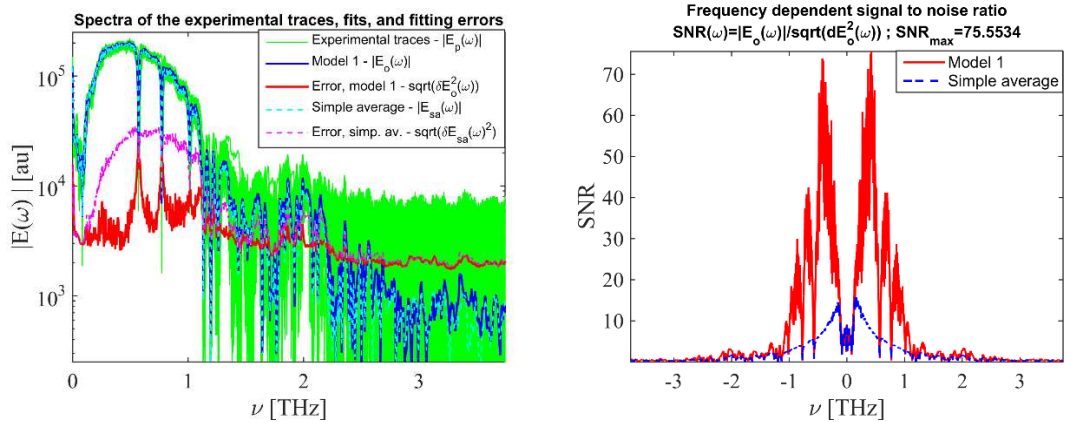

Figure 4. Comparison of the quality of the fit using model 1 versus simple average approximation. Left panel: spectra of the experimental traces, those of various fits and the corresponding fitting errors. Right panel: frequency dependent signal to nose ratio for various fits.

In Fig. 5 we present some of the statistical properties of traces plotted versus the trace number. As THz traces are acquired consecutively at equal time intervals (every 4 min), this statistical data essentially presents variation of the trace properties as a function of time over the period of one day. For example, Fig. 5 (top left) presents time shifts $\delta t_p$ in the THz pulse temporal position over time as fitted using model 1. From the figure, we note "fast" trace-to-trace variation in the value of time shifts on a scale of $\sim 0.04 \, ps$. This variation is directly related to the absolute position repeatability of the mechanical stage used in the optical delay line which is specked at $\delta x \gtrsim 1.5 \, \mu m$, which in turn, corresponds to the uncertainty in the trace position of $\delta t \sim 2\delta x/c \gtrsim 0.01 ps$. Moreover, we observe an overall "slow" drift of the THz pulse position over time that can be as much as $0.2 \, ps$ after all 400 measurement. Next, Fig. 5 (top right) shows changes in the trace power over time, where trace power is defined to be proportional to the square of the gain coefficient. From the figure, we note that "fast" trace-to-trace power variation is quite small and is on the order of $\sim 1 \, \%$. At the same time, we also note a "slow" variation of the average value of trace power, which can be as high as 10-20% over the time scale of the whole experiment (one day). The underlying reason for such strong power variation is not quite clear as the measurements of the laser power (see insert in Fig. 5 (top right)) show less than 1 % variation during the whole experiment (insert: power meter 1), while antenna aging effects usually lead to reduction of the THz power over time and not to the power increase as seen in Fig. 5. Most probable cause of such power variation is in the micro-misalignment of various optical elements during the experiment, particularly the ones installed on the moving delay line. An indirect confirmation of this is seen in the insert (power meter 2) in Fig. 5 top right. There, we show laser intensity variation after the delay line when a portion of the laser beam is diverted and focused onto the power meter through a $100 \, \mu m$ aperture. This simulates variation in the power of the laser beam that is focused onto the detector antenna after an optical delay line.

Strong power variations of ~10% were observed over the course of the measurements making us to suspect that it is the accumulation of the small trace-to-trace changes in the optical alignment or optical path quality related to the optical delay line which are responsible for large "slow" variation in the trace optical power. Another potential cause for the long-term variations in the trace power (beyond antenna aging) can be spontaneous loss of the mode locking in the laser with a consequent recovery in a somewhat different mode locking configuration. Although this power variation mechanism was indeed observed in several of our measurements, it is typically characterized by a much dramatic and sudden change in the trace power (30-60%) compared to 10-20% relatively slow power change that is observed in majority of our measurements.

Moreover, we note that the average trace power typically stays relatively constant over 10's or even 100's of traces, while suddenly changing to another value over a span of several measurements, which is again a testament of some permanent changes either in the optical path or the laser beam quality during the course of the experiment. We can say that in terms of power, traces can be subdivided into different clusters, each one characterised by an almost constant average power and relatively small power variation within each cluster. This clustering of the traces can be also seen in Fig. 5 bottom left and right panels where we show trace phase and frequency averaged trace error variations. Thus, in Fig. 5 (bottom left) we show trace phase defined as a phase of the complex gain coefficient $arg(C_p)$, and observe that it generally varies in a relatively smooth manner from trace to trace, while exhibiting sudden jumps between different clusters of traces. The trace clustering phenomenon is particularly evident when plotting the frequency averaged fit error for each trace $\delta E_p$ versus the trace gain coefficient $|C_p|$ (see Fig. 5 bottom right panel). The frequency averaged fit error for a trace $p$ is defined as:

$$\delta E_p^2 = \frac{1}{N_\omega} \sum_{n=1}^{N_\omega} \left| \delta E_p(\omega_n) \right|^2. \quad (20)$$

Finally, as we have discussed earlier, Fig 5 bottom right panel can, in theory, provide justification for the choice of a particular noise model (7), (8) or (9) that reflects best the nature of the "fast" noise in a pump-probe setup. For example, if the noise is independent of the trace power (model 1), then according to (7) we should expect that $\delta E_p$ is also independent of $|C_p|$; at the same time, using model 2 (8), or model 3 (9) to describe the same trace set, will result in $\delta E_p \sim 1/|C_p|$ dependence. Unfortunately, when analysing experimental data, frequently (however, not always) we cannot see statistically significant difference between the results of different models, due to large standard deviation of the noise amplitude. To demonstrate this point, in the insert in the Fig. 5 bottom right panel we show results of using model 2 and observe very similar behaviour of the noise amplitude versus the trace gain coefficient as in the case of model 1. Further studies are, therefore, necessary to address this issue, which would entail comparison between the models under different power excitation/detection and environmental conditions, which is beyond the scope of this paper.

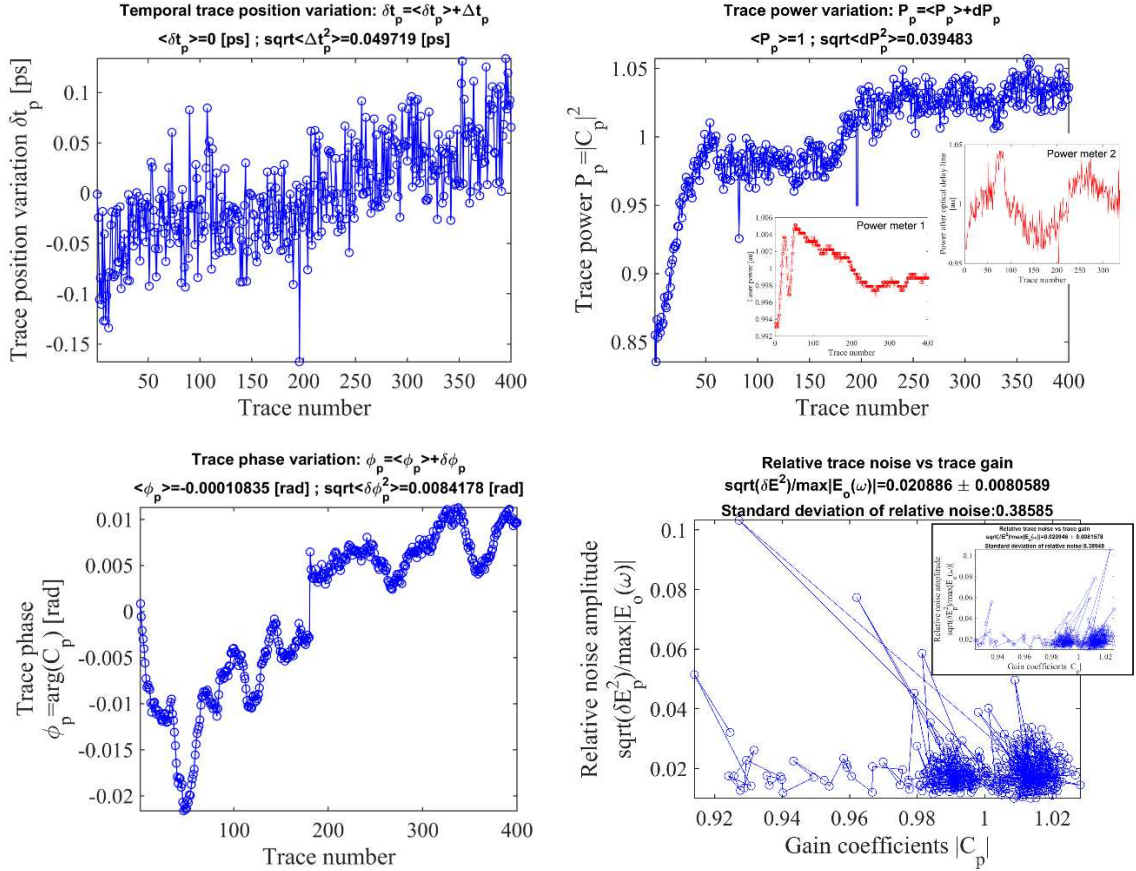

Figure 5. Statistical analysis of 400 THz traces using model 1. Top left panel: trace position shift vs the trace number. Top right: trace power vs the trace number; insert 1 – laser power before the optical delay line measured before acquisition of each THz pulse; insert 2 – laser power after the optical delay line measured before acquisition of each THz pulse. Bottom left: trace phase vs the trace number. Bottom right: frequency averaged trace noise vs the trace number; insert – same data but obtained using model 2.

## 7. Discussion of phase unwrapping and Model 3

Here we discuss some fundamental problems when trying to use model 3 in data fitting. We remind the readers that if we suppose that the noise is proportional both to the pulse amplitude $C_p$ and the nominal pulse intensity $E_o$, then we can write the pulse and its spectrum as in (9):

$$E_p(t) = C_p E_o(t - \delta t_p)[1 + \delta_p(t - \delta t_p)]$$

$$E_p(\omega) = C_p E_o(\omega) e^{-i\omega \delta t_p}[1 + \delta_p(\omega)]. \quad (21)$$

where $\delta_p(\omega)$ is the relative noise term for the pulse $p$. Taking natural logarithm of the pulse spectrum, and assuming that the amplitude of the relative noise is small $|\delta_p(\omega)| \ll 1$, we can rewrite (21) as:

$$\delta_p(\omega) = ln\big(E_p(\omega)\big) - ln\big(C_p\big) - ln\big(E_o(\omega)\big) + i\omega\delta t_p. \quad (22)$$

Here, it is important to mention a fundamental problem that arises when trying to compute a natural logarithm of a complex physical property. In particular, complex spectrum of a physical pulse can be represented using its frequency dependent absolute value and phase as $E_p(\omega) = |E_p(\omega)|e^{i\varphi_p(\omega)}$. For a physical property, like a pulse spectrum, its phase is typically a continuous function of frequency. However, when computing numerically natural logarithm of a complex, frequency dependant function one gets the value of a "wrapped" phase confined to the $[0,2\pi)$ interval that has a "saw" pattern when plotted versus frequency:

$$ln(E(\omega)) = ln(|E(\omega)|) + i \cdot mod(\varphi(\omega), 2\pi). \quad (23)$$

Therefore, equation (22) cannot be used directly when formulating an optimization problem as it assumes that natural logarithms (used in (22)) result in true continuous phases of the functions rather than the folded ones. Therefore, in order to use (22), one has to first "unwrap" a phase found numerically using (23). An operation of phase unwrapping typically starts with a phase value $\varphi_1$ computed using (23) at the edge $\omega_1$ of the frequency interval of interest. Then, one considers the value of the phase $\varphi_2$ computed using (23) at the adjacent frequency $\omega_2$. One assumes that the frequency grid is dense enough so that the phase change from one frequency to another is slow so that normally $|\varphi_2 - \varphi_1| < tol_\varphi \ll 2\pi$, where $tol_\varphi$ is a certain tolerance parameter. This condition, however, is broken when continuous phase $\varphi(\omega)$ reaches $|\varphi_1|{\sim}2\pi$ at $\omega_1$ and goes over $2\pi$ at $\omega_2$, which results in $|\varphi_2|{\sim}0$ due to phase wrapping by (23). The phase "unwrapping" algorithms thus start at the edge of a frequency interval of interest, and then go from one frequency to another, find the $2\pi$ phase jumps caused by (23), and correct them by adding an appropriate $2\pi$ multiplier so that the resultant "unwrapped" phase $\varphi(\omega)$ is a continuous function in a sense that for any two adjacent frequencies $|\varphi(\omega_2) - \varphi(\omega_1)| < tol_\varphi$. We also note that even if the phase unwrapping is perfectly implemented, the resulting phase is still different from the true one $\varphi(\omega)$ by a constant $2\pi n$ additive factor, where $n$ is some integer. This is because the value of the first phase at the edge of the frequency interval $\varphi_1$ can only be found up to a constant $2\pi n$ additive factor.

Yet another complication is that the realistic pulse spectra normally show strong absorption regions (water vapour absorption lines, for example), where the pulse spectral intensity can become lower than the noise level, and, as a consequence, the phase is scrambled. Therefore, when performing phase unwrapping, one normally finds several ($N_u$) disjoint frequency regions of successful phase unwrapping. We denote such frequency intervals as $W_r$, where $r = 1, \ldots, N_u$. We then denote as $N_\omega^r$ the number of frequencies within each interval $W_r$.

We now modify the fitting function (22) in order to account for the unknown $2\pi n$ additive factors that arise during phase unwrapping. Particularly, we define a phase correction function $\varphi_{p,r}$ that takes $N_t \times N_u$ values and that modifies (23) as:

$$\delta_p(\omega) = ln\big(E_p(\omega)\big) - ln\big(C_p\big) - ln\big(E_o(\omega)\big) + i\omega\delta t_p + i\varphi_{p,r}; \; \omega \in W_r. \quad (24)$$

Expression (23) means that within each interval of successful phase unwrapping $W_r$ we have to correct for the $2\pi n_{p,r}$ shifts that were introduced when performing phase unwrapping of different traces. Similarly, in place of (22), we can write a new definition of the fitting function that is more suitable when working with logarithms of the complex functions:

$$E_p(\omega) = C_p E_o(\omega) e^{-i\omega\delta t_p - i\varphi_{p,r}}\big[1 + \delta_p(\omega)\big]; \; \omega \in W_r. \quad (25)$$

## 8. Application of Model 3 to analysis of the THz time traces

In this section, we analyse the same 400 THz time traces as presented earlier, but now using model 3. First step is to unwrap the frequency dependent phase $Im\big(ln\big(E_p(\omega)\big)\big)$ of each of the traces, as well as to find the common frequency intervals $W_r$ (shared by all the traces) of successful phase unwrapping. In Fig. 6 we present unwrapped phases of all the traces (top panel), as well as a chart (bottom panel) that identifies positions of the common frequency intervals of successful phase unwrapping and a number of frequencies in each of such intervals. During phase unwrapping we require that each frequency interval of successful phase unwrapping contains at least $min(N_\omega^r) = 60$ frequency points. From Fig. 6 we observe that there are $r = 6$ of such intervals. In principle, we can set $min(N_\omega^r)$ to its lowest possible value of 2, however, the intervals containing a small number of frequencies will typically have a significant phase noise contribution, which would affect negatively the quality of the fit. Our experience tell us that the minimal value of frequencies in each interval $min(N_\omega^r)$ should be greater than 10-20 to guarantee that the results of the fit using model 3 are consistent with those obtained using models 1 and 2. As it is evident from Fig. 6, due to the use of a complex logarithm function, the unwrapped phases of different traces feature distinct $2\pi n_{p,r}$ additive factors within each of the intervals of successful phase unwrapping. These phase shifts $\varphi_{p,r}$ (p=1,…, 400 ; r=1,…,6) can be recovered and compensated for by using a more advanced form of the fitting function (25).

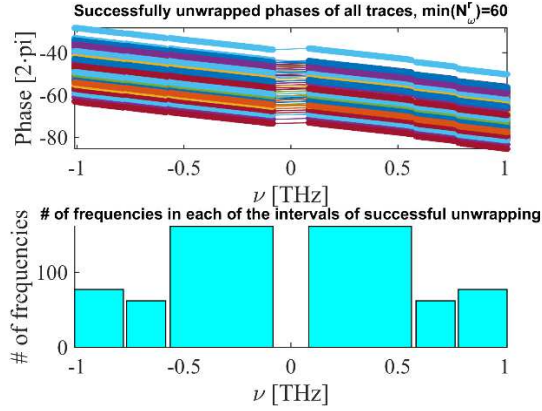

Figure 6. Top panel: unwrapped phases of all the traces. Bottom panel: a chart identifying positions of the common frequency intervals of successful phase unwrapping and a number of frequencies in each of the intervals.

Thus, in Fig. 7 left panel we show the same unwrapped phases as in Fig. 6, however compensated with the appropriate $\varphi_{p,r}$ phase correction factors $Im\left(ln\left(E_p(\omega)\right)\right) - \varphi_{p,r}; \; \omega \in W_r$. Clearly, unwrapped phases of all the traces collapse into the same curve. While, we expect that all the $\varphi_{p,r}$ factors should be proportional to $2\pi$, in practice we find that this only holds approximately. Thus, in Fig. 7 right panel we plot $mod\left(\varphi_{p,r}, 2\pi\right)$ and discover that while their absolute values are relatively small ($< 0.4 \; rad$), however, they are not zero.

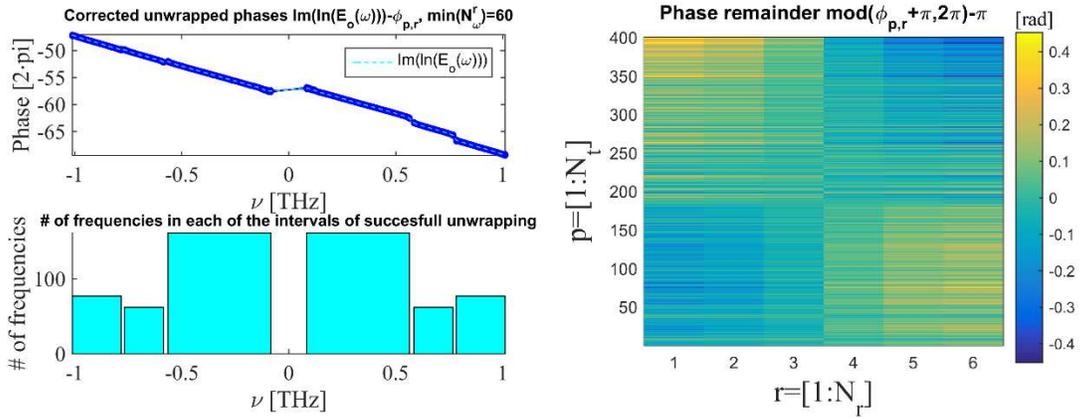

Figure 7. Left panel: unwrapped phases (as in Fig. 6) compensated with the appropriate $\varphi_{p,r}$ phase correction factors. Right panel: $mod\left(\varphi_{p,r}, 2\pi\right)$ which is generally expected to be close to zero.

The reason for such a behaviour (as detailed in Appendix 3) is in the competition between the two phase factors $\omega \delta t_p$ and $\varphi_{p,r}$ that appear in the model 3 (see Eq. (24)), and it is in some sense an artefact of this model. This artefact, furthermore, makes challenging a direct comparison of the values of the trace shifts $\delta t_p$ as predicted by models 1, 2 versus model 3. We illustrate this problem by plotting in Fig. 8 top left panel the time shifts of all the traces as calculated using model 3 (solid lines), as well as time shifts (dotted lines) as computed by model 1 (model 2 gives time shifts virtually identical to those of model 1). We see that because of the non-zeros $mod(\varphi_{p,r}, 2\pi)$ contributions, the value of the time shifts predicted by the two models are somewhat different. For the rest of the statistical parameters such as trace power and trace phase we observe an overall excellent correspondence between model 1 and model 3 (see Figs. 8 top right and bottom left panels), however trace-to-trace variation of these parameters are more significant in the case of model 3. This is especially pronounced when comparing the frequency averaged trace noise for the two models (see Fig. 8 bottom right), which is almost 50% larger for model 3 than for model 1. For model 3, the frequency averaged fit error for trace $p$ is defined as:

$$\delta E_p^2 = \frac{1}{N_u} \sum_{r=1}^{N_u} \frac{1}{N_\omega^r} \sum_{n=1}^{N_\omega^r} \left| E_o(\omega_n^r) \delta_p(\omega_n^r) \right|^2. \quad (26)$$

In fact, higher level of noise associated with model 3 compared to model 1 is easy to explain by noting that the number of frequencies used in fitting within model 3 is significantly lower than that used within model 1. This is because in model 3 we use only the frequencies within the common intervals of successful phase unwrapping, which are confined to <1THz in this particular example (see Fig. 6). At the same time, model 1 uses all the frequencies up to 3.75 THz. As the nose amplitude decreases with the number of frequencies used in the fit, it is not surprising that the results given by model 3 are noisier than those given by model 1.

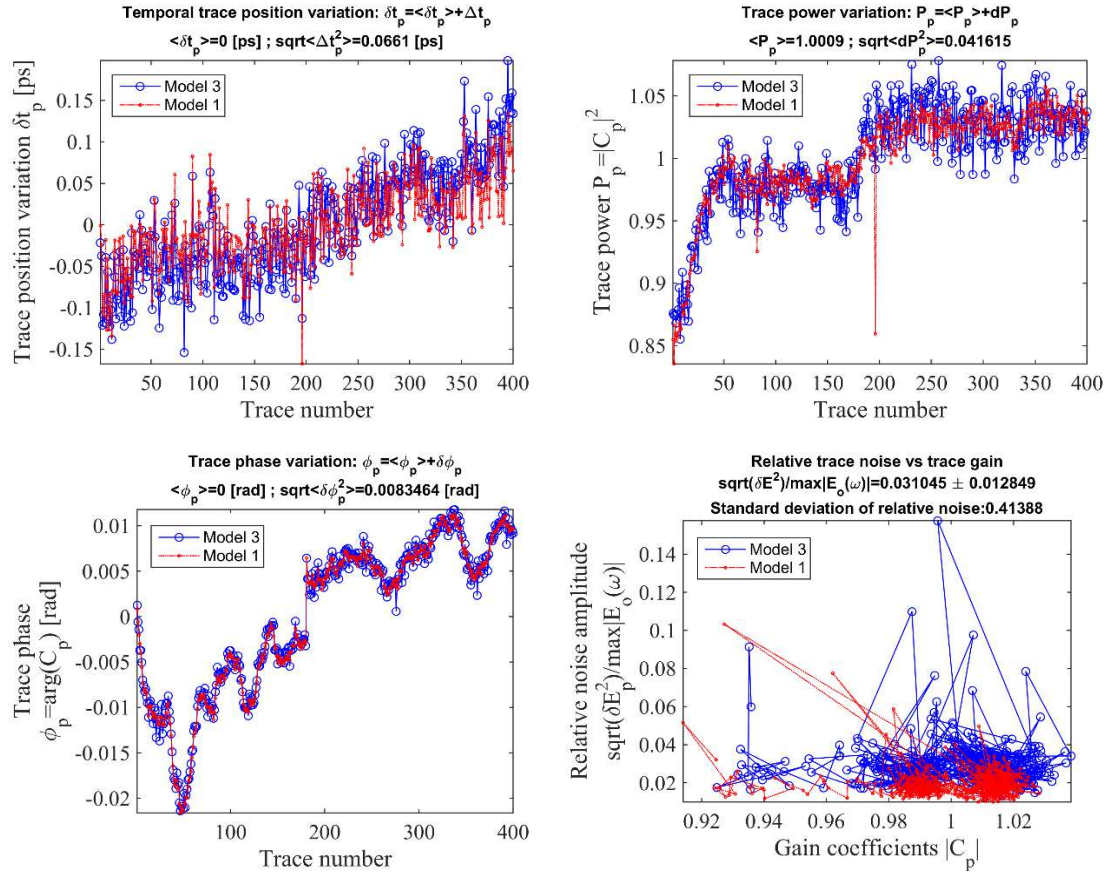

Figure 8. Statistical analysis of 400 THz traces using model 3 (solid blue), and comparison with model 1 (dashed red). Top left panel: trace position shift vs the trace number. Top right: trace power vs the trace number. Bottom left: trace phase vs the trace number. Bottom right: frequency averaged trace noise vs the trace number.

Finally, in Fig. 9 (left panel) we plot spectra $E_p(\omega)$ of the experimental pulses (solid green), spectrum of the fitted nominal pulse $E_o(\omega)$ (solid blue), and a spectrum of the fitting error $\delta E_o(\omega)$ (solid red) defined as a trace-wise average of the individual relative fitting errors $\delta_p(\omega)$ (18) multiplied by the nominal pulse spectrum:

$$\delta E_o^2(\omega) = \frac{1}{N_t}\sum_{p=1}^{N_t}\left|E_o(\omega)\delta_p(\omega)\right|^2. \quad (27)$$

Similarly to the results of model 1, we see that the fitting error resulting from model 3 is much smaller than that of the error associated with a simple average approximation. This is especially evident when plotting a frequency dependent signal to noise ratio (Fig. 9 right panel) defined as in (19). Finally, we note that SNR for the fit using model 3 is almost twice as small as that using model 1, which is consistent with the earlier discussion about the noisier nature of model 3.

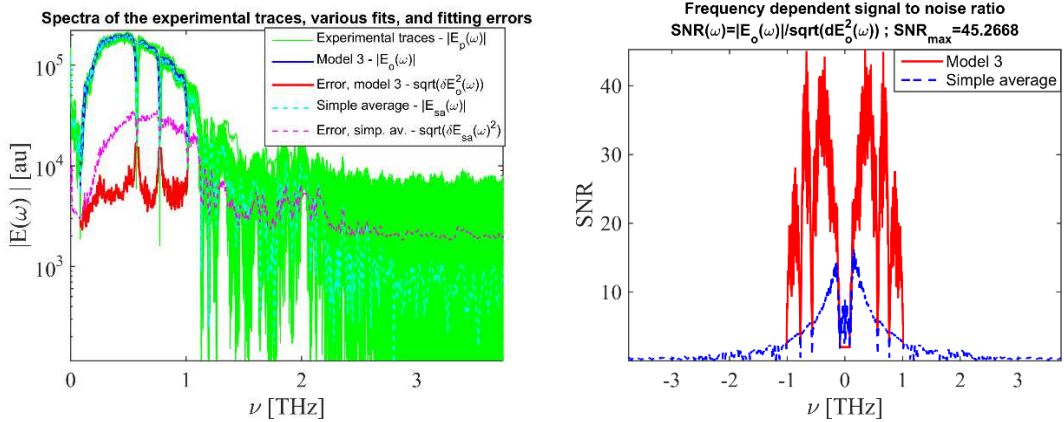

Figure 9. Comparison of the quality of the fit using model 3 versus a simple average approximation. Left panel: spectra of the experimental traces, those of various fits and the corresponding fitting errors. Right panel: frequency dependent signal to nose ration for various fits.

## 9. Conclusion

In this paper, we have first discussed the main types of noise in a typical pump-probe system, and then focused specifically on a THz-TDS setup. Next, we introduced three advanced fitting models for the pulses obtained in such systems, and detailed rigorous mathematical algorithms to find the corresponding fitting parameters. Finally, we performed a comparative analysis of the performance, advantages and limitations of the three algorithms by testing them on the experimental data collected using a particular TDS-THz system available in our laboratories.

More specifically, in our models we distinguish the "fast" noise, which is considered to be effectively negated by a lock-in amplifier, from the "slow" noise that describes trace-to-trace variations in the pulse power, pulse phase, and pulse spectral position (jitter). While the three fitting models include in the same way effects of the "slow" noise, however, they are different in the way they treat the leftover "fast" noise, which is not completely compensated by a lock-in amplifier. These differences result in three distinct mathematical algorithms that are presented in great details and that feature very different sets of computational advantages and limitations.

Overall, we observe that models 1 and 2 are superior to model 3 in terms of the resultant fitting errors, the overall quality of the fit, and numerical stability. This is related to the fact that model 3 only uses data at frequencies at which phase unwrapping is successful for all the traces, while models 1,2 use all the frequencies in the pulse spectra. Moreover, even if the phase unwrapping is successful, at the edges of the spectrum, noise contribution to the phases can be significant so that computation of various averages using $mod(\dots, 2\pi)$ operation can become unstable. This leads to further limitation on the choice of frequencies that can be used in the fitting procedure and renders model 3 to be somewhat unpredictable in terms of the quality of a fit.

A clear advantage of the model 3 compared to the models 1,2 is that it employs only a linear system of equations that can be solved analytically in a closed form. Therefore, computational effort associated with model 3 is significantly smaller than that associated with models 1,2 which require solution of a system of nonlinear equations.

Finally, another important strength of the model 3 is that it gives a properly unwrapped phase of a nominal trace, which can be further used for statistical analysis of signals in more complex algorithms like a cut-back method.

## 10. Appendix 1. Model 1, mathematical formulation

In this section, we use the following form of the fitting function to fit the THz trace spectra:

$$E_p(\omega) = \left( C_p E_o(\omega) + \delta E_p(\omega) \right) e^{-i\omega_n \delta t_p}, \quad (10.1)$$

from which we also obtain:

$$\delta E_p(\omega) = E_p(\omega) e^{i\omega_n \delta t_p} - C_p E_o(\omega). \quad (10.2)$$

Here we note that the analytical form of the fitting function (10.2) also implies certain ambiguity in the values of the gain coefficients $C_p$ and the nominal field $E_o$. In particular, after performing the following substitutions $C_p \rightarrow C_p/C_o$, $E_o \rightarrow C_o E_o$, where $C_o$ is any complex number, equations (10.2) will remain unchanged. This ambiguity can be lifted by making certain assumptions about the physical nature of the registered pulses. For example, the complex value of $C_o$ can be found by assuming that the total energy of the nominal pulse is equal to the average energy of all the registered pulses, while the phase of the nominal pulse equals to the average phase of all the pulses. Moreover, from (10.2) it is also clear that the time shifts are defined only up to an arbitrary constant $\delta t_o$. Thus, after performing the following substitutions $\delta t_p \rightarrow \delta t_p + \delta t_o$, $E_o(t) \rightarrow E_o(t + \delta t_o)$ equations (10.2) will remain unchanged. In turn, this ambiguity can be resolved by assuming, for example, that the average of the time shifts $\delta t_p$ is zero, which physically means that the nominal pulse $E_o(t)$ is positioned in the "middle" between all the measured pulses $E_p(t)$. Finally, we note that anytime there is an ambiguity in the values of some coefficients in the analytical form of the fitting function, it will have an important impact on the mathematical formulation of the problem as such ambiguities result in the underdetermined (or degenerate) system of equations. This can be resolved by assuming certain normalisation or constraint conditions, that should be chose using certain physical reasoning about the system.

### 10.1 Formulation of the optimization problem

In order to find the complex gain factors $C_p$, time shifts $\delta t_p$, and a spectrum of the nominal THz field $E_o(\omega)$ we define an optimization problem with respect to the spectrally and trace-wise averaged value of the noise. In other words, we look for the values of the abovementioned parameters and functions that minimize the following weighting function $Q$:

$$Q = \frac{1}{N_t} \sum_{p=1}^{N_t} \frac{1}{N_\omega} \sum_{n=1}^{N_\omega} \left| \delta E_p(\omega_n) \right|^2 = \frac{1}{N_t} \sum_{p=1}^{N_t} \frac{1}{N_\omega} \sum_{n=1}^{N_\omega} \left| E_p^n e^{i\omega_n \delta t_p} - C_p E_o^n \right|^2, \quad (10.3)$$

where $N_t$ is the number of THz traces used in fitting, $N_\omega$ is the number of frequency components in the Fourier spectrum of pulses, and for the sake of brevity we define $E_o^n = E_o(\omega_n)$, $E_p^n =$ $E_p(\omega_n)$. To find the minimum of the weighting function we have to solve a system of $N_t + N_\omega$ linear equations:

$$\frac{\partial Q}{\partial E_o^{n*}} = 0; \; n = 1 \dots N_\omega, \quad (10.4)$$

$$\frac{\partial Q}{\partial C_p^*} = 0; \; p = 1 \dots N_t, \quad (10.5)$$

and $N_t$ nonlinear equations:

$$\frac{\partial Q}{\partial \delta t_p} = 0; \; p = 1 \dots N_t. \quad (10.6)$$

## 10.2 Optimal spectra of the nominal pulse

Remembering that

$$Q = \frac{1}{N_t} \sum_{p=1}^{N_t} \frac{1}{N_\omega} \sum_{n=1}^{N_\omega} \left( E_p^n e^{i\omega_n \delta t_p} - C_p E_o^n \right) \left( E_p^{n*} e^{-i\omega_n \delta t_p} - C_p^* E_o^{n*} \right), \quad (10.7)$$

Equation (10.4) becomes:

$$\frac{1}{N_t} \sum_{p=1}^{N_t} \left( E_p^n e^{i\omega_n \delta t_p} - C_p E_o^n \right) C_p^* = 0, \quad (10.8)$$

from which we get an expression for the optimal spectra of the nominal trace:

$$E_o^n = \frac{\frac{1}{N_t} \sum_{p=1}^{N_t} C_p^* E_p^n e^{i\omega_n \delta t_p}}{\langle P \rangle}, \quad (10.9)$$

where the average power relative to that of a nominal trace is defined as $\langle P \rangle = \frac{1}{N_t} \sum_{p=1}^{N_t} |C_p|^2$.

## 10.3 Gain coefficients and their normalization

Substituting (10.9) into (10.7), and after simplifications we get:

$$Q = \langle |E|^2 \rangle - \frac{1}{N_t^2} \frac{1}{\langle P \rangle} \sum_{p'p''} C_{p'}^* C_{p''} \langle E_{p'} E_{p''}^* e^{i\omega_n (\delta t_{p'} - \delta t_{p''})} \rangle, \quad (10.10)$$

where frequency and trace averaged power $\langle |E|^2 \rangle = \frac{1}{N_t} \sum_{p=1}^{N_t} \frac{1}{N_\omega} \sum_{n=1}^{N_\omega} |E_p^n|^2$, and

$\langle E_{p'} E_{p''}^* e^{i\omega_n (\delta t_{p'} - \delta t_{p''})} \rangle = \frac{1}{N_\omega} \sum_{n=1}^{N_\omega} E_{p'}^n E_{p''}^{n*} e^{i\omega_n (\delta t_{p'} - \delta t_{p''})}$. Eq. (10.10) can also be written in the matrix form as:

$$Q = \langle |E|^2 \rangle - \frac{\overline{C}^+ M \overline{C}}{\overline{C}^+ \overline{C}}, \quad (10.11)$$

where $\overline{C} = \left(C_1, \dots C_{N_t}\right)^T$ is a vector of the complex gain coefficients, $\overline{C}^+$ is a conjugate transpose of $\overline{C}$, while M is a Hermitian matrix with elements:

$$M_{p'p''} = \frac{1}{N_t} \langle E_{p'} E_{p''}^* e^{i\omega_n \left(\delta t_{p'} - \delta t_{p''}\right)} \rangle. \quad (10.12)$$

In what follows we assume that values of the time shifts $\delta t_p$ are known, which will allow us to solve minimization problem (10.11) using linear algebra. In particular, we minimize (10.11) using a constraint on the norm of the vector of the gain coefficients $\frac{1}{N_t} \overline{C}^+ \overline{C} = \langle P \rangle$, where $\langle P \rangle$ is considered constant. For example, by choosing $\langle P \rangle = 1$ we impose that the power carried by the nominal pulse $E_o$ equals to the average power carried by all the experimentally measured pulses $E_p$. Minimization of (10.11) with constraint is equivalent to minimization of the following weighting function:

$$Q = \langle |E|^2 \rangle - \frac{1}{\langle P \rangle} \frac{1}{N_t} \overline{C}^+ M \overline{C} + \lambda \left(\frac{1}{N_t} \overline{C}^+ \overline{C} - \langle P \rangle\right), \quad (10.13)$$

where $\lambda$ is an additional optimization variable and $\langle P \rangle$ is a constant. Minimization of (10.13) requires that:

$$\frac{\partial Q}{\partial \lambda} = 0 \implies \frac{1}{N_t} \overline{C}^+ \overline{C} = \langle P \rangle, \quad (10.14)$$

$$\frac{\partial Q}{\partial \overline{C}^+} = 0 \implies \frac{1}{\langle P \rangle} M \overline{C} = \lambda \overline{C}, \quad (10.15)$$

Equation (10.14) represents the power normalization condition, while equation (10.15) is an eigen value problem with respect to the vector of the gain coefficients. Thus, by choosing $\lambda = \lambda_{max}/\langle P \rangle$ and $\overline{C} = \overline{C}_{\lambda_{max}}$, where $\lambda_{max}$ and $\overline{C}_{\lambda_{max}}$ are the largest positive eigen value and the corresponding eigenvector of the Hermitian matrix M, and by further normalizing the vector $\overline{C}$ to respect normalization condition (10.14), we solve the minimization problem (10.11) and find:

$$Q = \langle |E|^2 \rangle - \lambda_{max}, \quad (10.16)$$

Note that due to the functional form (10.12) of the elements of matrix M, the eigen value $\lambda_{max}$ is only dependent on the temporal shifts $\delta t_p$.

## 10.4 Choosing the common phase factor for the gain coefficients

Finally, we need to address the remaining ambiguity in the definition of the vector of gain coefficients. Particularly, by multiplying it by any complex phase factor $\overline{C} \longrightarrow e^{i\varphi_0}\overline{C}$ the value of the weighting function (10.16) will not change. In order to determine the value of the phase $\varphi_0$, we have to make an additional assumption on the phase relation between the nominal pulse $E_o$ and the measured pulses $E_p$. For example, by assuming that there is only a minor change in the phases between $E_o$ and the measured pulses $E_p$, we can demand that the vector of the gain coefficients should be as real as possible. Thus, $\varphi_0$ can be found by minimising, for example, the following weighting function:

$$q \; = \sum_{p=1}^{N_t} \frac{Im(e^{i\varphi_0} C_p)^2}{|C_p|^2}, \quad (10.17)$$

After certain simplifications, (10.17) can be rewritten as:

$$q = \sum_{p=1}^{N_t} \frac{Im(C_p)^2}{|C_p|^2} + \sum_{p=1}^{N_t} \frac{\left(Re(C_p)^2 - Im(C_p)^2\right) \sin {\varphi_0}^2 + Re(C_p)\, Im(C_p) \sin 2\varphi_0}{|C_p|^2}. \quad (10.18)$$

In order to minimize q in (10.18), we then solve the following equation:

$$\frac{\partial q}{\partial \varphi_0} = 0 \longrightarrow \varphi_0 = \frac{\pi}{2} m - \frac{1}{2} tan^{-1} \left( \frac{2 \sum_{p=1}^{N_t} \frac{Re(C_p)\, Im(C_p)}{|C_p|^2}}{\sum_{p=1}^{N_t} \frac{Re(C_p)^2 - Im(C_p)^2}{|C_p|^2}} \right), \quad (10.19)$$

where the integer m $= [1,2,3,4]$ has to be chosen to minimize the value of $q$ in (10.18). Finally, the optimal vector of the gain coefficients have to be replaced by the $\overline{C} \longrightarrow e^{i\varphi_0} \overline{C}$.

## 10.5 Temporal shifts of the traces

So far, we have assumed that the temporal time shifts $\delta t_p$ are known. This allowed us to solve analytically equation (10.4) and find expression (10.9) for the spectra of the nominal pulse $E_o^n$. We then used linear algebra to show that the gain coefficients $C_p$ can be found by solving a standard eigen value problem (10.16), and we found that the original weighting function (10.3) have been transformed into (10.15), which is dependent only on the vector of the time shifts $\overline{\delta t} = \left[\delta t_1, \ldots, \delta t_{N_t}\right]^T$:

$$Q\left(\overline{\delta t}\right) = \langle |E|^2 \rangle - \lambda_{max}(\delta \bar t). \quad (10.20)$$

Direct minimization of Q as defined in (10.20) will not result in a unique solution with respect to $\overline{\delta t}$ as time shifts $\delta t_p$ are known only up to an arbitrary constant $t_o$. Indeed, if $\overline{\delta t}$ is a solution of $\min \left( Q\left(\overline{\delta t}\right) \right)$, then $\overline{\delta t} + \delta t_o$ is also a solution. This is a consequence of the fact that eigen value $\lambda_{max}$ in (10.20) is that of a matrix M with elements defined by (10.12). There, all the elements depend only on the pair-wise differences in the time shifts $\delta t_{p'} - \delta t_{p''}$, which means that addition of the same constant to all the individual time shifts will not change the value of the weighting function (10.20). Therefore, in order to guarantee a unique solution of the minimization problem (10.20) we have to impose an additional constraint on the values of the time shifts. One such possible constraint can be in terms of the average value of the time shifts $\frac{1}{N_t} \sum_{p=1}^{N_t} \delta t_p = \delta t_o$, which in the vector form can be written as:

$$\mathbf{1}^T . \overline{\delta t} = N_t \delta t_o, \quad (10.21)$$

where $\mathbf{1}^T = (1, \dots, 1)$ is an identity vector of length $N_t$. In the absence of additional information it is logical to assume that $\delta t_o = 0$, meaning that the nominal trace $E_o(t)$ should be found somewhere in the "middle" between the measured traces $E_p(t)$. In the presence of constraint (10.21), the weighting function should be thus modified as:

$$Q(\overline{\delta t}) = \langle |E|^2 \rangle - \lambda_{max}(\overline{\delta t}) - \lambda \left( \mathbf{1}^T . \overline{\delta t} - N_t \delta t_o \right), \quad (10.22)$$

where $\lambda$ is a new optimization variable. Minimization of (10.22) requires solution of the normalization equation and $N_t$ nonlinear equations (10.6), which in the vector form can be written in terms of a gradient of the weighting function with respect to the vector of the time shifts:

$$\frac{\partial Q}{\partial \lambda} = 0 \implies \mathbf{1}^T \overline{\delta t} = N_t \delta t_o, \quad (10.23)$$

$$\bar{\nabla} Q(\overline{\delta t}) = 0 \implies -\bar{\nabla} \lambda_{max}(\overline{\delta t}) - \lambda \mathbf{1} = 0. \quad (10.24)$$

A standard nonlinear iterative Newton method can be employed to find a numerical solution of (10.24). Particularly, assuming that after k iterations $\overline{\delta t_k}$ is an approximation to the vector of time shifts, in order to find a better approximation $\overline{\delta t_{k+1}} = \overline{\delta t_k} + \overline{\Delta t_{k+1}}$ during iteration $k+1$ one solves the following equation:

$$\bar{\nabla} Q(\overline{\delta t_k} + \overline{\Delta t_{k+1}}) = -\bar{\nabla} \lambda_{max}(\overline{\delta t_k} + \overline{\Delta t_{k+1}}) - \lambda \mathbf{1} = 0, \quad (10.25)$$

where $\overline{\Delta t_{k+1}}$ is a correction to the approximation $\overline{\delta t_k}$. Using Taylor expansions of (10.25) we can write in the matrix form:

$$\bar{\nabla} Q(\overline{\delta t_k} + \overline{\Delta t_{k+1}}) = -\bar{\nabla} \lambda_{max}(\overline{\delta t_k}) - H(\overline{\delta t_k}) \overline{\Delta t_{k+1}} - \lambda \mathbf{1} + O\left( \left( \overline{\Delta t_{k+1}} \right)^2 \right) = 0, \quad (10.26)$$

where $H(\overline{\delta t})$ is the Hessian matrix of $\lambda_{max}(\overline{\delta t})$. By retaining only the linear terms in (10.26) we can find the correction $\overline{\Delta t_{k+1}}$ to the vector of the time shifts as:

$$\overline{\Delta t_{k+1}} = -H(\overline{\delta t_k})^{-1} \left( \bar{\nabla} \lambda_{max}(\overline{\delta t_k}) + \lambda \mathbf{1} \right). \quad (10.27)$$

Finally, in order to satisfy constraint (10.23) for the new time shifts $\overline{\delta t}_{k+1} = \overline{\delta t}_k + \overline{\Delta t}_{k+1}$, and assuming that $\overline{\delta t}_k$ already respects the constraint (10.23), we have to choose $\lambda$ in (10.27) so that $\mathbf{1}^T \overline{\Delta t}_{k+1} = 0$, which leads to the following expression for $\lambda$:

$$\lambda = -\frac{\mathbf{1}^T H(\overline{\delta t}_k)^{-1} \overline{\nabla} \lambda_{max}(\overline{\delta t}_k)}{\mathbf{1}^T H(\overline{\delta t}_k)^{-1} \mathbf{1}}. \quad (10.28)$$

In the expressions (10.27, 10.28), the elements of the gradient vector $\overline{\nabla} \lambda_{max}$ and the Hessian matrix $H$ are computed numerically using the following definitions, as well as the standard finite difference expressions for the first and second order derivatives:

$$\overline{\nabla} \lambda_{max}(\overline{\delta t}_k) = \left[\frac{\partial \lambda_{max}(\overline{\delta t})}{\partial \delta t_1}, \dots, \frac{\partial \lambda_{max}(\overline{\delta t})}{\partial \delta t_{N_t}}\right]^T \Bigg|_{\overline{\delta t} = \overline{\delta t}_k}, (10.25)$$

$$H(\delta \bar{t}_k) = \begin{bmatrix} \frac{\partial^2 \lambda_{max}(\overline{\delta t})}{\partial \delta t_1^2} & \cdots & \frac{\partial^2 \lambda_{max}(\overline{\delta t})}{\partial \delta t_1 \partial \delta t_{N_t}} \\ \cdots & \cdots & \cdots \\ \frac{\partial^2 \lambda_{max}(\overline{\delta t})}{\partial \delta t_{N_t} \partial \delta t_1} & \cdots & \frac{\partial^2 \lambda_{max}(\overline{\delta t})}{\partial \delta t_{N_t}^2} \end{bmatrix} \Bigg|_{\overline{\delta t} = \overline{\delta t}_k}, (10.25)$$

where we remind the reader that $\lambda_{max}(\overline{\delta t})$ is the largest positive eigen value of the Hermitian matrix $M$ defined by its elements (10.12). In practice, calculation of the $N_t^2$ elements of the full Hessian matrix can be time intensive, so one normally retains only the $N_t$ diagonal elements of the Hessian matrix to speed up the calculations. As an initial guess for the time shifts (required to start the Newton algorithm) one can simply take $\overline{\delta t}_1 = \mathbf{1}^T \delta t_o$. With this choice of the initial guess the Newton method will normally converge rapidly in only a few iterations.

## 11. Appendix 2. Model 2, mathematical formulation

In this section we use the following form of the fitting function to fit the THz trace spectra:

$$E_p(\omega) = \frac{1}{a_p}\Big(E_o(\omega) + \delta E_p(\omega)\Big)e^{-i\omega_n\delta t_p}, \quad (11.2)$$

where $a_p = \frac{1}{c_p}$, and the noise spectrum is:

$$\delta E_p(\omega) = a_p E_p(\omega)e^{i\omega_n\delta t_p} - E_o(\omega). \quad (11.3)$$

He we note that the analytical form of the fitting function (11.3) also implies certain ambiguity in the values of the expansion coefficients $a_p$ and the nominal field $E_o$. In particular, after performing the following substitutions $a_p \rightarrow a_o a_p$, $E_o \rightarrow a_o E_o$, $\delta E_p \rightarrow a_o \delta E_p$, where $a_o$ is any complex number, equations (11.3) will remain unchanged. This ambiguity can be lifted by making certain assumptions about the physical nature of the registered pulses. For example, the complex value of $a_o$ can be found by assuming that the total energy of the nominal pulse is equal to the average energy of all the registered pulses, while the phase of the nominal pulse equals to the average phase of all the pulses. Finally, as it is the case in model 1, from (11.2) it is also clear that the time shifts are defined only up to an arbitrary constant $\delta t_o$. In turn, this ambiguity can be resolved by assuming, for example, that the average of the time shifts $\delta t_p$ is zero.

### 11.1 Formulation of the optimization problem

In order to find the complex factors $a_p$, time shifts $\delta t_p$, and a spectrum of the nominal THz field $E_o(\omega)$ we define an optimization problem with respect to the spectrally and trace-wise averaged value of the noise. In other words, we look for the values of the abovementioned parameters and functions that minimize the following weighting function $Q$:

$$Q = \frac{1}{N_t}\sum_{p=1}^{N_t}\frac{1}{N_\omega}\sum_{n=1}^{N_\omega}\big|\delta E_p(\omega_n)\big|^2 = \frac{1}{N_t}\sum_{p=1}^{N_t}\frac{1}{N_\omega}\sum_{n=1}^{N_\omega}|a_p E_p^n e^{i\omega_n\delta t_p} - E_o^n|^2, \quad (11.4)$$

where as before, $N_t$ is the number of THz traces used in fitting, $N_\omega$ is the number of frequency components in the pulse Fourier spectra, and for the sake of brevity we define $E_o^n = E_o(\omega_n)$, $E_p^n = E_p(\omega_n)$. To find the minimum of the weighting function we have to solve a system of $N_t + N_\omega$ linear equations:

$$\frac{\partial Q}{\partial E_o^{n*}} = 0; \; n = 1 \dots N_\omega, \quad (11.5)$$

$$\frac{\partial Q}{\partial a_p^*} = 0; \; p = 1 \dots N_t, \quad (11.6)$$

and $N_t$ nonlinear equations:

$$\frac{\partial Q}{\partial \delta t_p} = 0; \ p = 1 \ldots N_t. \quad (11.7)$$

## 11.2 Optimal spectra of the nominal pulse

Remembering that:

$$Q = \frac{1}{N_t} \sum_{p=1}^{N_t} \frac{1}{N_\omega} \sum_{n=1}^{N_\omega} (E_o^n - a_p E_p^n e^{i\omega_n \delta t_p})(E_o^{n*} - a_p^* E_p^{n*} e^{-i\omega_n \delta t_p}), \quad (11.8)$$

equation (11.5) becomes:

$$\frac{1}{N_t} \sum_{p=1}^{N_t} (E_o^n - a_p E_p^n e^{i\omega_n \delta t_p}) = 0, \quad (11.9)$$

from which we get an expression for the optimal spectra of the nominal trace:

$$E_o^n = \frac{1}{N_t} \sum_{p=1}^{N_t} a_p E_p^n e^{i\omega_n \delta t_p}. \quad (11.10)$$

## 11.3 Expansion coefficients and their normalization

Substituting (11.10) into (11.8), and after simplifications we get:

$$Q = \frac{1}{2N_\omega} \sum_{n=1}^{N_\omega} \frac{1}{N_t^2} \sum_{p'p''=1}^{N_t} \left| a_{p'} E_{p'}^n e^{i\omega_n \delta t_{p'}} - a_{p''} E_{p''}^n e^{i\omega_n \delta t_{p''}} \right|^2. \quad (11.11)$$

One can also demonstrate that (11.11) can be written in the matrix form as follows:

$$Q = \overline{a}^+ M \overline{a}, \quad (11.12)$$

where $\overline{a} = (a_1, \ldots a_{N_t})^T$, $\overline{a}^+$ is a conjugate transpose of $\overline{a}$, and Hermitian matrix $M$ is:

$$M = \frac{1}{N_t^2} \begin{bmatrix} (N_t - 1)\langle |E_1|^2 \rangle & -\langle E_2 E_1^* \rangle & \ldots & -\langle E_{N_t} E_1^* \rangle \\ -\langle E_1 E_2^* \rangle & (N_t - 1)\langle |E_2|^2 \rangle & \ldots & \vdots \\ \vdots & \vdots & \ddots & \vdots \\ -\langle E_1 E_{N_t}^* \rangle & -\langle E_2 E_{N_t}^* \rangle & \ldots & (N_t - 1)\langle |E_{N_t}|^2 \rangle \end{bmatrix}, \quad (11.13)$$

where $\langle E_{p'} E_{p''}^* \rangle = \frac{1}{N_\omega} \sum_{n=1}^{N_\omega} E_{p'}^n E_{p''}^{n*} e^{i\omega_n (\delta t_{p'} - \delta t_{p''})}$.

In what follows we assume that the values of the time shifts $\delta t_p$ are known, which will allow us to solve minimization problem (11.11) using linear algebra. In order to minimize (11.11) we have to solve a system (11.6) of $N_t$ linear equations. We note, however that the linear equations are degenerate as $a_p = 0$ constitutes a trivial solution that minimizes (11.11). We therefore, have to minimize (11.11) using a certain constraint on the norm of the vector of the expansion coefficients. One of the possible choices is to ensure that the power of the nominal pulse equals to the average power of the traces. Remembering the meaning of the expansion coefficients $a_p$ as a ratio of the amplitude of the nominal pulse to that of the measured trace $a_p = \frac{1}{c_p} \sim \frac{E_o}{E_p}$ we can then write the abovementioned normalisation condition as:

$$\frac{1}{N_t} \sum_{p=1}^{N_t} |a_p|^2 \langle |E_p|^2 \rangle = \langle |E|^2 \rangle, \quad (11.14)$$

where frequency and trace averaged power is $\langle |E|^2 \rangle = \frac{1}{N_t} \sum_{p=1}^{N_t} \frac{1}{N_\omega} \sum_{n=1}^{N_\omega} |E_p^n|^2$, while the frequency averaged power of trace $p$ is $\langle |E_p|^2 \rangle = \frac{1}{N_\omega} \sum_{n=1}^{N_\omega} |E_p^n|^2$. In matrix form (14) can be written as:

$$\overline{a}^+ D \overline{a} = \langle |E|^2 \rangle, \quad (11.15)$$

where diagonal matrix $D$ has the following elements $D_{p,p} = \frac{1}{N_t} \langle |E_p|^2 \rangle$.

With an additional constraint (11.14), minimization of (11.11) is equivalent to minimization of the following weighting function:

$$Q = \overline{a}^+ M \overline{a} - \lambda \left( \overline{a}^+ D \overline{a} - \langle |E|^2 \rangle \right), \quad (11.16)$$

where $\lambda$ is an additional optimization variable. Minimization of (11.16) requires that:

$$\frac{\partial Q}{\partial \lambda} = 0 \implies \overline{a}^+ D \overline{a} = \langle |E|^2 \rangle, \quad (11.17)$$

$$\frac{\partial Q}{\partial \overline{a}} = 0 \implies M \overline{a} = \lambda D \overline{a}. \quad (11.18)$$

Equation (11.17) is a normalization condition, while equation (11.18) presents a generalized linear eigen value problem which is solved by choosing $\lambda$ to be the eigen value and $\overline{a} = \overline{a}_\lambda$ to be a corresponding eigen vector of the matrix $D^{-1}M$. Furthermore, $\overline{a}_\lambda$ has to be normalized to respect (11.17). We can then rewrite weighting function (11.16) as simply:

$$Q = \lambda \langle |E|^2 \rangle \Rightarrow min(Q) = \lambda_{min} \langle |E|^2 \rangle, \quad (11.19)$$

from which we conclude that in order to minimize weighting function (11.11) we have to choose $\overline{a} = \overline{a}_{\lambda_{min}}$ to be the eigen vector that corresponds to the smallest positive eigen value $\lambda_{min}$ of the Hermitian matrix $D^{-1}M$. Furthermore, $\overline{a}_\lambda$ has to be normalized to respect (11.17).

## 11.4 Choosing the common phase factor for the expansion coefficients

Finally, from the form of the weighting function (11.11) and the normalisation condition (11.15) we note that even an optimal choice of $\overline{a}$ is still defined up to an arbitrary phase multiplier $e^{i\varphi_0}$. In order to determine the value of the phase $\varphi_0$, we have to make an additional assumption on the phase relation between the nominal pulse $E_o$ and the measured pulses $E_p$. For example, similarly to the considerations of model 1, we can assume that there is only a minor change in the phases between $E_o$ and the measured pulses $E_p$. Thus, we can demand that the vector of the expansion coefficients should be as real as possible. Thus, $\varphi_0$ can be found by minimizing the following weighting function (see model 1):

$$q = \sum_{p=1}^{N_t} \frac{Im(e^{i\varphi_0}a_p)^2}{|a_p|^2}. \quad (11.20)$$

## 11.5 Temporal shifts of the traces

So far, we have assumed that the temporal time shifts $\delta t_p$ are known. This allowed us to solve analytically equation (11.5) and find expression (11.10) for the spectra of the nominal pulse $E_o^n$. We then used linear algebra to show that the expansion coefficients $a_p$ can be found by solving a generalised eigen value problem (11.18), and we found that the original weighting function (11.4) have been transformed into (11.19), which is dependent only on the vector of the temporal time shifts $\overline{\delta t}$:

$$Q\left(\overline{\delta t}\right) = \lambda_{min}(\delta \overline{t})\langle |E|^2\rangle. \quad (11.21)$$

Similarly to model 1, direct minimization of $Q$ as defined in (11.21) will not result in a unique solution with respect to $\overline{\delta t}$ as time shifts $\delta t_p$ are known only up to an arbitrary constant $t_o$. Therefore, in order to guarantee a unique solution of the minimization problem (11.21) we impose an additional constraint on the average value of the time shifts:

$$\mathbf{1}^T. \overline{\delta t} = N_t \delta t_o, \quad (11.22)$$

In the presence of a constraint, the weighting function should be thus modified as:

$$Q\left(\overline{\delta t}\right) = \lambda_{min}(\delta \overline{t})\langle |E|^2\rangle - \lambda\left(\mathbf{1}^T. \overline{\delta t} - N_t \delta t_o\right), \quad (11.23)$$

where $\lambda$ is a new optimization variaable. Minimization of the weighting function with a constraint (11.23) requires solution of the normalization equation and $N_t$ nonlinear equations (11.7), which in the vector form can be written in terms of a gradient of the weighting function with respect to the vector of the time shifts:

$$\frac{\partial Q}{\partial \lambda} = 0 \implies \mathbf{1}^T \overline{\delta t} = N_t \delta t_o, \quad (11.24)$$

$$\bar{\nabla} Q\left(\overline{\delta t}\right) = 0 \implies \bar{\nabla} \lambda_{min}\left(\overline{\delta t}\right) \langle |E|^2 \rangle - \lambda \mathbf{1} = 0. \quad (11.25)$$

Similarly to model 1, a standard nonlinear iterative Newton method can be employed to find a numerical solution of (11.25). Particularly, assuming that after $k$ iterations $\overline{\delta t}_k$ is an approximation to the vector of time shifts, in order to find a better approximation $\overline{\delta t}_{k+1} = \overline{\delta t}_k + \overline{\Delta t}_{k+1}$ during iteration $k + 1$ one solves the following equation:

$$\bar{\nabla} Q\left(\overline{\delta t}_k + \overline{\Delta t}_{k+1}\right) = \bar{\nabla} \lambda_{min}\left(\overline{\delta t}_k + \overline{\Delta t}_{k+1}\right) \langle |E|^2 \rangle - \lambda \mathbf{1} = 0, \quad (11.26)$$

where $\overline{\Delta t}_{k+1}$ is a correction to the approximation $\overline{\delta t}_k$. Using Taylor expansions of (11.26) we can write in the matrix form:

$$\bar{\nabla} Q\left(\overline{\delta t}_k + \overline{\Delta t}_{k+1}\right) = \left(\bar{\nabla} \lambda_{min}\left(\overline{\delta t}_k\right) + H\left(\overline{\delta t}_k\right)\overline{\Delta t}_{k+1}\right) \langle |E|^2 \rangle - \lambda \mathbf{1} + O\left(\left(\overline{\Delta t}_{k+1}\right)^2\right) = 0,$$
(11.27)

where $H\left(\overline{\delta t}\right)$ is the Hessian matrix of $\lambda_{min}\left(\overline{\delta t}\right)$. By retaining only the linear terms in (11.27) we can find the correction $\overline{\Delta t}_{k+1}$ to the vector of the time shifts as:

$$\overline{\Delta t}_{k+1} = -H\left(\overline{\delta t}_k\right)^{-1} \left(\bar{\nabla} \lambda_{min}\left(\overline{\delta t}_k\right) - \frac{\lambda}{\langle |E|^2 \rangle} \mathbf{1}\right). \quad (11.28)$$

Finally, in order to satisfy constraint (11.24) for the new time shifts $\overline{\delta t}_{k+1} = \overline{\delta t}_k + \overline{\Delta t}_{k+1}$, and assuming that $\overline{\delta t}_k$ already respects the constraint (11.24), we have to choose $\lambda$ in (11.28) so that $\mathbf{1}^T \overline{\Delta t}_{k+1} = 0$, which leads to the following expression for $\lambda$:

$$\lambda = \frac{\mathbf{1}^T H\left(\overline{\delta t}_k\right)^{-1} \bar{\nabla} \lambda_{min}\left(\overline{\delta t}_k\right)}{\mathbf{1}^T H\left(\overline{\delta t}_k\right)^{-1} \mathbf{1}} \langle |E|^2 \rangle. \quad (11.28)$$

where we remind the reader that $\lambda_{min}\left(\overline{\delta t}\right)$ is the smallest eigen value of the Hermitian matrix $M$ defined in (11.13). In practice, calculation of the $N_t^2$ elements of the full Hessian matrix can be time intensive, so one normally retains only the $N_t$ diagonal elements of the Hessian matrix to speed up the calculations. As an initial guess for the time shifts (required to start the Newton algorithm) one can simply take $\overline{\delta t}_1 = \mathbf{1}^T \delta t_o$. With this choice of the initial guess the Newton method will normally converge rapidly in only a few iterations.

## 12. Appendix 3. Model 3, mathematical formulation

In this section we use the following form of the fitting function to fit the THz trace spectra:

$$E_p(\omega) = C_p E_o(\omega) e^{-i\omega \delta t_p - i\varphi_{p,r}} \big[ 1 + \delta_p(\omega) \big]; \ \omega \in W_r. \quad (12.5)$$

where $\delta_p(\omega)$ is the relative noise term for the pulse $p$ while $W_r$, $r = 1, \dots, N_u$ denote $N_u$ intervals of successful phase unwrapping of the THz pulses. Taking natural logarithm of the pulse spectrum, and assuming that the amplitude of the relative noise is small $|\delta_p(\omega)| \ll 1$, we can write:

$$\delta_p(\omega) = ln\big(E_p(\omega)\big) - ln\big(C_p\big) - ln\big(E_o(\omega)\big) + i\omega \delta t_p + i\varphi_{p,r}; \ \omega \in W_r. \quad (12.4)$$

Here we note that the analytical form of the fitting functions (12.4) also implies certain ambiguity in the values of the logarithms of the gain coefficients $ln(C_p)$, nominal field $ln(E_o(\omega))$, and the correction phase $\varphi_{p,r}$. In particular, after performing the following substitutions $ln(C_p) \rightarrow ln(C_p) + ln(C_o)$, $ln(E_o(\omega)) \rightarrow ln(E_o(\omega)) - ln(C_o)$, where $C_o$ is any complex number, equations (12.4) will remain unchanged. This ambiguity can be lifted by making certain assumptions about the physical nature of the registered pulses. For example, the complex value of $C_o$ can be found by assuming that the amplitude and phase of the nominal pulse should be similar to those of the measured pulses, which can be expressed via a normalization condition $\sum_p ln(C_p) = 0$. Additionally, if we fix $r$, and after performing the following substitutions $\varphi_{p,r} \rightarrow \varphi_{p,r} + \varphi_r$; $\forall p$, as well as $ln(E_o(\omega)) \rightarrow ln(E_o(\omega)) - i\varphi_r$; $\omega \in W_r$, where $\varphi_r$ is any phase value, equations (12.4) will remain unchanged. In order to remove this ambiguity in the value of the correction phase we can request, for example, the following normalization conditions $\sum_{p=1}^{N_t} \varphi_{p,r} = 0$; $\forall r$. Similarly, if we fix $p$, after performing the following substitutions $\varphi_{p,r} \rightarrow \varphi_{p,r} + \varphi_p$; $\forall r$, as well as $ln(C_p) \rightarrow ln(C_p) + i\varphi_p$, where $\varphi_p$ is any phase value, equations (12.4) will remain unchanged. In order to lift this ambiguity in the value of the correction phase we can impose, for example, the following normalization conditions $\sum_{r=1}^{N_u} \varphi_{p,r} = 0$; $\forall p$. Finally, similarly to the discussions of models 1, 2, we note that the value of the time shifts in (12.4) are defined only up to an arbitrary constant $\delta t_o$. Indeed, after performing the following substitutions $\delta t_p \rightarrow \delta t_p + \delta t_o$, $ln(E_p(\omega)) \rightarrow ln(E_p(\omega)) + i\omega \delta t_p$ (which is identical to $E_o(t) \rightarrow E_o(t + \delta t_o)$) equation (12.4) will remain unchanged. In turn, this ambiguity can be resolved by assuming, for example, that the average of the time shifts $\delta t_p$ is zero in order to guarantee that the nominal pulse is situated in the "middle" between the measured pulses.

## 12.1 Formulation of the optimization problem

In order to find logarithms of the complex gain factors $ln(C_p)$, logarithms of the spectrum of the nominal pulse $ln(E_o(\omega))$, as well as time shifts $\delta t_p$ we define an optimization problem with respect to the spectrally and trace-wise averaged value of the relative noise. In other words, we look for the values of the abovementioned parameters and functions that minimize a certain weighting function $Q$, namely:

$$Q = \frac{1}{N_t} \sum_{p=1}^{N_t} \frac{1}{N_u} \sum_{r=1}^{N_u} \frac{1}{N_\omega^r} \sum_{n=1}^{N_\omega^r} \left| \delta_p(\omega_r^n) \right|^2 =$$

$$= \frac{1}{N_t} \sum_{p=1}^{N_t} \frac{1}{N_u} \sum_{r=1}^{N_u} \frac{1}{N_\omega^r} \sum_{n=1}^{N_\omega^r} \left| lE_{p,r}^n - lC_p - lE_{o,r}^n + i\omega_r^n \delta t_p + i\varphi_{p,r} \right|^2, \qquad (12.6)$$

where, again, $N_t$ is the number of THz traces used in fitting, $N_u$ is the number of disjoint frequency regions of successful phase unwrapping $W_r$ ($r = 1, \dots, N_u$), $N_\omega^r$ is the number of frequencies within each interval $W_r$, and finally $\omega_r^n$ are the actual frequencies ($n = 1, \dots, N_\omega^r$) in each of the intervals $W_r$. Furthermore, for the sake of brevity we define $lE_{o,r}^n = ln(E_o(\omega_r^n))$, $lE_{p,r}^n = ln\left(E_p(\omega_r^n)\right)$. To find the minimum of the weighting function we have to solve a system of the following linear equations:

$$\frac{\partial Q}{\partial lE_{o,r}^{n*}} = 0; \; r = 1, \dots, N_u; \; n = 1 \dots N_\omega^r, \quad (12.7)$$

$$\frac{\partial Q}{\partial lC_p^*} = 0; \; p = 1 \dots N_t, \quad (12.8)$$

$$\frac{\partial Q}{\partial \delta t_p} = 0; \; p = 1 \dots N_t, \quad (12.9)$$

$$\frac{\partial Q}{\partial \varphi_{p,r}} = 0; \; p = 1, \dots, N_t; \; r = 1, \dots, N_u. \quad (12.10)$$

## 12.2 Optimal spectra of the nominal pulse

Using definition of the weighting function (12.6), equation (12.7) then becomes:

$$\frac{\partial Q}{\partial lE_{o,r}^{n*}} = 0 \; \Rightarrow \; \frac{1}{N_t} \sum_{p=1}^{N_t} \left( lE_{p,r}^n - lC_p - lE_{o,r}^n + i\omega_r^n \delta t_p + i\varphi_{p,r} \right) = 0. \quad (12.11)$$

We can furthermore simplify the equation above by using various normalization conditions:

$$\frac{1}{N_t} \sum_{p=1}^{N_t} lC_p = i\langle \varphi_C \rangle, \quad (12.12)$$

$\sum_{p=1}^{N_t} \delta t_p = 0,$    (12.13)

$\frac{1}{N_t}\sum_{p=1}^{N_t} \varphi_{p,r} = \langle \varphi_r \rangle,$    (12.14)

where $\langle \varphi_C \rangle$ and $\langle \varphi_r \rangle$ denote certain fixed values of the phase averages that we will specify later. Using definitions (12.12-12.14) we get the following expression for the optimal spectra of the nominal trace:

$lE_{o,r}^n = \frac{1}{N_t}\sum_{p=1}^{N_t} lE_{p,r}^n + i(\langle \varphi_r \rangle - \langle \varphi_C \rangle).$    (12.15)

## 12.3 Gain coefficients, time shifts, and phase correction term

Now that the spectra of the nominal pulse (12.16) is known, we proceed with finding the gain coefficients using equation (12.8):

$\frac{\partial Q}{\partial lC_p^*} = 0 \implies \frac{1}{N_u}\sum_{r=1}^{N_u} \frac{1}{N_\omega^r}\sum_{n=1}^{N_\omega^r}\left( lE_{p,r}^n - lC_p - lE_{o,r}^n + i\omega_r^n \delta t_p + i\varphi_{p,r} \right) = 0.$    (12.16)

We can furthermore simplify the equation above by defining:

$\frac{1}{N_u}\sum_{r=1}^{N_u} \varphi_{p,r} = \langle \varphi_p \rangle,$    (12.17)

where $\langle \varphi_p \rangle$ denote certain fixed values of the phase averages that we will specify later. We then simplify (12.16) and find:

$lC_p = \langle lE_p - lE_o \rangle + i\langle \omega \rangle \delta t_p + i\langle \varphi_p \rangle,$    (12.18)

where we have defined the following averages over all the frequencies:

$\langle \omega \rangle = \frac{1}{N_u}\sum_{r=1}^{N_u} \frac{1}{N_\omega^r}\sum_{n=1}^{N_\omega^r} \omega_r^n,$    (12.19)

$\langle \omega^2 \rangle = \frac{1}{N_u}\sum_{r=1}^{N_u} \frac{1}{N_\omega^r}\sum_{n=1}^{N_\omega^r} (\omega_r^n)^2,$    (12.20)

$\langle lE_p - lE_o \rangle = \frac{1}{N_u}\sum_{r=1}^{N_u} \frac{1}{N_\omega^r}\sum_{n=1}^{N_\omega^r}\left( lE_{p,r}^n - lE_{o,r}^n \right),$    (12.21)

$\langle \omega\left( lE_p - lE_o \right) \rangle = \frac{1}{N_u}\sum_{r=1}^{N_u} \frac{1}{N_\omega^r}\sum_{n=1}^{N_\omega^r} \omega_r^n\left( lE_{p,r}^n - lE_{o,r}^n \right).$    (12.22)

From equation (12.18) it follows that the gain coefficients are related to the time shifts, therefore equations (12.8) and (12.9) have to be solved simultaneously. Using equation (12.9) we find:

$\frac{\partial Q}{\partial \delta t_p} = 0 \Rightarrow \frac{1}{N_u} \sum_{r=1}^{N_u} \frac{1}{N_\omega^r} \sum_{n=1}^{N_\omega^r} \omega_r^n \left( Im \left( lE_{p,r}^n - lE_{o,r}^n \right) - Im \left( lC_p \right) + \omega_r^n \delta t_p + \varphi_{p,r} \right) = 0.$ (12.23)

Defining the following averages across the individual intervals of successful unwrapping $W_r$:

$\langle \omega \rangle_r = \frac{1}{N_\omega^r} \sum_{n=1}^{N_\omega^r} \omega_r^n,$    (12.24)

$\langle lE_p - lE_o \rangle_r = \frac{1}{N_\omega^r} \sum_{n=1}^{N_\omega^r} \left( lE_{p,r}^n - lE_{o,r}^n \right),$    (12.25)

we can rewrite (12.23) as:

$\langle \omega^2 \rangle \delta t_p = \langle \omega \rangle Im \left( lC_p \right) - Im \langle \omega \left( lE_p - lE_o \right) \rangle - \frac{1}{N_u} \sum_{r=1}^{N_u} \langle \omega \rangle_r \varphi_{p,r}.$    (12.27)

From the form of equation (12.27) it follows that the gain coefficients and the time shifts are also related to the phase correction terms, therefore equations (12.8), (12.9) and (12.10) have to be solved simultaneously. Using equation (12.10) we find:

$\frac{\partial Q}{\partial \varphi_{p,r}} = 0 \Rightarrow \frac{1}{N_\omega^r} \sum_{n=1}^{N_\omega^r} \left( Im \left( lE_{p,r}^n - lE_{o,r}^n \right) - Im \left( lC_p \right) + \omega_r^n \delta t_p + \varphi_{p,r} \right) = 0.$    (12.28)

Using definitions of various averages (12.24, 12.25), we can rewrite (12.28) as:

$\varphi_{p,r} = Im \left( lC_p \right) - Im \langle lE_p - lE_o \rangle_r - \langle \omega \rangle_r \delta t_p.$    (12.29)

Now we can solve analytically a system of coupled linear equations (12.18), (12.27), (12.29) to find the final expressions for the nominal pulse spectrum, gain coefficients, time shifts and phase correction terms:

$lE_{o,r}^n = \frac{1}{N_t} \sum_{p=1}^{N_t} lE_{p,r}^n + i \langle \varphi_r \rangle - i \langle \varphi_C \rangle,$    (12.30)

$\delta t_p = \frac{\frac{1}{N_u} \sum_{r=1}^{N_u} \langle \omega \rangle_r Im \langle lE_p - lE_o \rangle_r - Im \langle \omega \left( lE_p - lE_o \right) \rangle}{\langle \omega^2 \rangle - \frac{1}{N_u} \sum_{r=1}^{N_u} (\langle \omega \rangle_r)^2},$    (12.31)

note that $\delta t_p$ is independent of the choice of $\langle \varphi_r \rangle$, and $\langle \varphi_C \rangle$ phase factors, and it can be computed using $lE_{o,r}^n$ as given by (12.30) assuming, for example, that $\langle \varphi_r \rangle = \langle \varphi_C \rangle = 0$.

$Re \left( lC_p \right) = Re \langle lE_p - lE_o \rangle,$    (12.32)

$Im \left( lC_p \right) = Im \langle lE_p - lE_o \rangle + \langle \omega \rangle \delta t_p + \langle \varphi_p \rangle,$    (12.33)

$\varphi_{p,r} = Im \left( lC_p \right) - Im \langle lE_p - lE_o \rangle_r - \langle \omega \rangle_r \delta t_p,$    (12.34)

and there is also a self-consistency condition that follows from definitions (12.14) and (12.17):

$\frac{1}{N_t}\sum_{t=1}^{N_t}\langle\varphi_p\rangle = \frac{1}{N_u}\sum_{r=1}^{N_u}\langle\varphi_r\rangle = \frac{1}{N_t}\sum_{t=1}^{N_t}\frac{1}{N_u}\sum_{r=1}^{N_u}\varphi_{p,r}.$    (12.35)

## 12.4 Choosing the values of various phase averages

So far, we have solved analytically the problem of fitting a collection of traces and their spectra with the fitting functions of the form (12.4, 12.5). While solution to the minimization problem was found in the closed form (12.30-12.35), we still have to precise the values of various phase factors $\langle\varphi_r\rangle$, $\langle\varphi_p\rangle$, and $\langle\varphi_C\rangle$ in order to make this solution unique.

We start with the choice of $\langle\varphi_r\rangle$ factors. When choosing those factors, our goal is to guarantee that the inverse Fourier transform of the nominal $E_o(\omega)$ spectrum be similar to those of the original pulses. Consider in more details expression for the nominal pulse spectrum. Assuming that we deal with ideal pulses having spectra:

$E_p(\omega) = |C_p||E_o(\omega)|e^{i\varphi_{C_p}+i\varphi_o(\omega)-i\omega\delta t_p},$    (12.36)

where $\langle\varphi_C\rangle = \frac{1}{N_t}\sum_{p=1}^{N_t}\varphi_{C_p}$. After taking logarithm of (12.36) and unwrapping the phase we get:

$log\left(E_p(\omega)\right) = log(|C_p|) + log(|E_o(\omega)|) + i\varphi_{C_p} + i\varphi_o(\omega) - i\omega\delta t_p + i2\pi n_{p,r}; \ \omega \in W_r,$
(12.36)

or in a more compact notation:

$Im\left(lE_{p,r}^n\right) = \varphi_{C_p} + \varphi_o(\omega_r^n) - \omega_r^n\delta t_p + 2\pi n_{p,r},$    (12.37)

Where, as before, $W_r$ is a continuous frequency region of successful phase unwrapping, and $2\pi n_{p,r}$ are the distinct phase factors that appear for each of the traces during unwrapping procedure of the logarithm phases. Then, within each $W_r$ region we can write for the spectrum of the nominal pulse as given by (12.30) by using (12.37):

$Im\left(lE_{o,r}^n\right) = \frac{1}{N_t}\sum_{p=1}^{N_t}Im\left(lE_{p,r}^n\right) + \langle\varphi_r\rangle - \langle\varphi_C\rangle =$

$= \varphi_o(\omega_r^n) + \frac{2\pi}{N_t}\sum_{p=1}^{N_t}n_{p,r} + \langle\varphi_r\rangle.$    (12.38)

From (12.38) we see that because of the additional $2\pi n_{p,r}$ phase factors that appear for each of the traces during phase unwrapping procedure in each of the frequency intervals $W_r$, the phase of the nominal pulse will be somewhat shifted by $\frac{2\pi}{N_t}\sum_{p=1}^{N_t}n_{p,r}$, which is generally not commensurate with $2\pi$. In order to correct for this artefact and guarantee that $exp\left(i\cdot Im\left(lE_{o,r}^n\right)\right) = exp\left(i\cdot\varphi_o(\omega_r^n)\right)$, we compute $\langle\varphi_r\rangle$ as follows:

$$\langle \varphi_r \rangle = \frac{1}{N_t} \sum_{p=1}^{N_t} mod \left( \frac{1}{N_\omega^r} \sum_{n=1}^{N_\omega^r} \left( Im\left(lE_{p,r}^n\right) + \omega_r^n \delta t_p - \frac{1}{N_t} \sum_{p=1}^{N_t} Im\left(lE_{p,r}^n\right) \right), 2\pi \right) =$$

$$= \frac{1}{N_t} \sum_{p=1}^{N_t} mod \left( \varphi_{C_p} - \langle \varphi_C \rangle - \frac{1}{N_t} \sum_{p=1}^{N_t} 2\pi n_{p,r}, 2\pi \right) \approx -mod \left( \frac{2\pi}{N_t} \sum_{p=1}^{N_t} n_{p,r}, 2\pi \right). \quad (12.40)$$

The last approximation in (12.40) is valid if variation of phases of the gain coefficients is small, which is typically true in case of the high-quality experimental traces. With these definitions of $\langle \varphi_r \rangle$ the inverse Fourier transform of the nominal pulse spectrum as computed by (12.30) will look similar to the original pulses if we assume that $\langle \varphi_C \rangle = 0$.

From (12.33) and (12.37), while using (12.40) we get:

$$Im\left(lE_{o,r}^n\right) = \frac{1}{N_t} \sum_{p=1}^{N_t} Im\left(lE_{p,r}^n\right) + \langle \varphi_r \rangle - \langle \varphi_C \rangle \approx$$

$$\approx \varphi_o(\omega_r^n) + \frac{2\pi}{N_t} \sum_{p=1}^{N_t} n_{p,r} - mod \left( \frac{1}{N_t} \sum_{p=1}^{N_t} 2\pi n_{p,r}, 2\pi \right) =$$

$$\approx \varphi_o(\omega_r^n) + 2\pi m_r, \quad (12.41)$$

where $m_r$ is some integer that is different for different $W_r$ regions. Then, the gain coefficients become:

$$Im(lC_p) = Im\langle lE_p - lE_o \rangle + \langle \omega \rangle \delta t_p + \langle \varphi_p \rangle =$$

$$= \frac{1}{N_u} \sum_{r=1}^{N_u} \frac{1}{N_\omega^r} \sum_{n=1}^{N_\omega^r} \left( Im(lE_{p,r}^n) - Im(lE_{o,r}^n) + \omega_r^n \delta t_p \right) + \langle \varphi_p \rangle =$$

$$= \varphi_{C_p} + \frac{2\pi}{N_u} \sum_{r=1}^{N_u} (n_{p,r} - m_r) + \langle \varphi_p \rangle. \quad (12.42)$$

From (12.42) we see that because of the additional $2\pi(n_{p,r} - m_r)$ phase factors that appear for each of the traces during unwrapping procedure in each of the frequency intervals $W_r$, the phase of the gain coefficient for the pulse $p$ will be somewhat shifted by $\frac{2\pi}{N_u} \sum_{r=1}^{N_u} (n_{p,r} - m_r)$, which is generally not commensurate with $2\pi$. In order to correct for this artefact and guarantee that $Im(lC_p) = \varphi_{C_p}$, we have to choose $\langle \varphi_p \rangle = -\frac{2\pi}{N_u} \sum_{r=1}^{N_u} (n_{p,r} - m_r)$. In practice, direct calculation of the phase correction factors $\langle \varphi_p \rangle$ is challenging as $n_{p,r}$ are not known a priori. Instead, we compute them indirectly in two steps. First, we compute phases $Im(lC_p)$ of the complex gain factors using an expression similar to (12.33, 12.42):

$$Im(lC_p) = \frac{1}{N_u} \sum_{r=1}^{N_u} mod \left( Im\langle lE_p - lE_o \rangle_r + \langle \omega \rangle_r \delta t_p, 2\pi \right). \quad (12.43)$$

Using expressions (12.38) for $Im\left(lE_{p,r}^n\right)$ and (41) for $Im\left(lE_{o,r}^n\right)$ we can, indeed verify that definition (12.43) retrieves $mod\left(\varphi_{C_p}, 2\pi\right)$ for ideal traces of the form (12.36) as:

$$Im\left(lC_p\right) = \frac{1}{N_u}\sum_{r=1}^{N_u} mod\left(\frac{1}{N_\omega^r}\sum_{n=1}^{N_\omega^r}\left(Im\left(lE_{p,r}^n\right) - Im\left(lE_{o,r}^n\right) + \omega_r^n \delta t_p\right), 2\pi\right) =$$

$$= \frac{1}{N_u}\sum_{r=1}^{N_u} mod\left(\varphi_{C_p} + 2\pi\left(n_{p,r} - m_r\right), 2\pi\right) = mod\left(\varphi_{C_p}, 2\pi\right). \quad (12.44)$$

With $Im\left(lC_p\right)$ computed using (12.43), the phase correction factors $\langle\varphi_p\rangle$ are then defined using (33) as:

$$\langle\varphi_p\rangle = Im\left(lC_p\right) - Im\langle lE_p - lE_o \rangle - \langle\omega\rangle\delta t_p. \quad (12.45)$$

## 12.5 Advantages and limitations of model 3

Finally, we note that in the case of multiple regions of successful phase unwrapping, the form of the fitting function (12.5):

$$E_p(\omega) \sim C_p E_o(\omega) e^{-i\omega\delta t_p - i\varphi_{p,r}}; \ \omega \in W_r, \quad (12.46)$$

implies a certain phase variation $\varphi_{p,r}$ across the whole frequency spectrum. This phase change competes with the $\omega\delta t_p$ factor in (12.46) that is responsible for the time shift of a pulse after performing the inverse Fourier transform of $E_p(\omega)$. The contribution of $\varphi_{p,r}$ can, thus, lead to an additional time shift of the pulse, thus rendering the values $\delta t_p$ as computed using (12.31) to be somewhat unreliable. In this respect, utilisation of models 1,2 are more advantageous as they give much better approximation to the time shifts $\delta t_p$.

Additionally, model 3 uses only frequencies at which phase unwrapping is successful, while models 1,2 use all the frequencies. This leads to somewhat higher average errors in case of the model 3. Moreover, even if the phase unwrapping is successful, at the edges of the spectrum, noise contribution to the phases can be significant so that computation of various averages using $mod(\ldots, 2\pi)$ operation can become unstable. This leads to further limitation on the choice of frequencies that can be used in the fitting procedure and renders the method employed by the model 3 to be somewhat unpredictable in terms of the quality of a fit.

Similarly, if only a small number of traces are used in the fitting, the trace-to-trace noise variation, together with the use of $mod(\ldots, 2\pi)$ operation can render the algorithm of model 3 unstable. Therefore, model 3 performs best when a large number of traces are used.

A clear advantage of the model 3 compared to the models 1,2 is that it employs only a linear system of equations that can be solved analytically in a closed form. Therefore, model 3 is significantly faster than models 1,2 (that rely on solution of the nonlinear equations), especially in the case of a large number of traces.

Finally, the principal strength of the model 3 is that it gives a properly unwrapped phase (with removed artefacts) of a nominal trace, which can be further used for statistical analysis of signals in more complex algorithms like a cut-back method.